\PassOptionsToPackage{pdfpagelabels=false}{hyperref}
\documentclass[fleqn,usenatbib,useAMS]{mnras}
\usepackage{graphicx}
\usepackage{amssymb}
\usepackage{amsmath}
\usepackage[dvipsnames]{xcolor}
\usepackage[figure,figure*,table,table*]{hypcap}
\usepackage{subfig}
\usepackage{multirow}

\usepackage{ulem}

\newcommand{\hmpc}{h^{-1} \, \mathrm{Mpc}}
\newcommand{\hkpc}{h^{-1} \, \mathrm{kpc}}

\title[BOSS LOWZ simulation-based RSD analysis]{Five-percent measurements of the growth rate from simulation-based modelling of redshift-space clustering in BOSS LOWZ}
\author[J. U. Lange et al.]{Johannes U. Lange$^{1, 2}$\thanks{email: jolange@ucsc.edu}, Andrew P. Hearin$^3$, Alexie Leauthaud$^1$,\newauthor Frank C. van den Bosch$^4$, Hong Guo$^5$ and Joseph DeRose$^{6, 7}$\\
$^1$Department of Astronomy and Astrophysics, University of California, Santa Cruz, CA 95064, USA\\
$^2$Kavli Institute for Particle Astrophysics and Cosmology and Department of Physics, Stanford University, CA 94305, USA\\
$^3$Argonne National Laboratory, Argonne, IL 60439, USA\\
$^4$Department of Astronomy, Yale University, New Haven, CT 06511, USA\\
$^5$Key Laboratory for Research in Galaxies and Cosmology, Shanghai Astronomical Observatory, Shanghai 200030, China\\
$^6$Berkeley Center for Cosmological Physics, University of California, Berkeley, CA 94720, USA\\
$^7$Santa Cruz Institute for Particle Physics, University of California, Santa Cruz, CA 95064,
USA}
\pubyear{2020}

\begin{document}

\date{Accepted xxx. Received xxx}

\label{firstpage}
\pagerange{\pageref{firstpage}--\pageref{lastpage}}

\maketitle

\begin{abstract}
    We use a simulation-based modelling approach to analyse the anisotropic clustering of the BOSS LOWZ sample over the radial range $0.4 \, \hmpc$ to $63 \, \hmpc$, significantly extending what is possible with a purely analytic modelling framework. Our full-scale analysis yields constraints on the growth of structure that are a factor of two more stringent than any other study on large scales at similar redshifts. We infer $f \sigma_8 = 0.471 \pm 0.024$ at $z \approx 0.25$, and $f \sigma_8 = 0.431 \pm 0.025$  at $z \approx 0.40$; the corresponding $\Lambda$CDM predictions of the Planck CMB analysis are $0.470 \pm 0.006$ and $0.476 \pm 0.005$, respectively. Our results are thus consistent with Planck, but also follow the trend seen in previous low-redshift measurements of $f \sigma_8$ falling slightly below the $\Lambda$CDM+CMB prediction. We find that small and large radial scales yield mutually consistent values of $f \sigma_8$, but there are $1-2.5 \sigma$ hints of small scales ($< 10 \, \hmpc$) preferring lower values for $f \sigma_8$ relative to larger scales. We analyse the constraining power of the full range of radial scales, finding that most of the multipole information about $f\sigma_8$ is contained in the scales $2 \, \hmpc \lesssim s \lesssim 20 \, \hmpc$. Evidently, once the cosmological information of the quasi-to-nonlinear regime has been harvested, large-scale modes contain only modest additional information about structure growth. Finally, we compare predictions for the galaxy--galaxy lensing amplitude of the two samples against measurements from SDSS and assess the lensing-is-low effect in light of our findings.
\end{abstract}


\section{Introduction}

The standard cosmological model postulates that the evolution of the Universe can be described by the interaction of (dark) matter, radiation, and dark energy in the context of general relativity. In particular, the simplest model that can accurately describe most if not all observations today is the $\Lambda$ cold dark matter ($\Lambda$CDM) cosmological model. Although the standard model has been remarkably successful in explaining numerous observational data sets in isolation, the growing diversity of high-precision cosmological measurements offers an opportunity to conduct stringent self-consistency tests of $\Lambda$CDM.

A very powerful combination of observations are the cosmic microwave background (CMB) coupled with measurements of the low-redshift Universe, as these can be used to test $\Lambda$CDM predictions for the state of the Universe at very different times of its evolution. Observations of the CMB alone place tight constraints on $\Lambda$CDM parameters, enabling precise, testable predictions for the late-time evolution of the Universe. We can generically divide these predictions into two categories: the expansion history of the Universe, and the growth of large-scale structure. Much attention has been paid to the apparent tension between the late-time expansion rate inferred from the CMB in comparison to direct observations \citep[see e.g.][and references therein]{Verde2019_NatAs_3_891}. Similarly, there is mounting evidence that growth of structure predictions from the CMB do not match what is seen in observations \citep{Abbott2018_PhRvD_98_3526, Hikage2019_PASJ_71_43, Asgari2020_arXiv_2007_5633}. Confirming and quantifying the level of disagreement in both aspects of the prediction is crucial to narrowing down theoretical explanations for $\Lambda$CDM tensions \citep{Blinov2019_PhRvL_123_1102, Vattis2019_PhRvD_99_1302, Keeley2019_JCAP_12_035, Ivanov2020_PhRvD_102_3502, Kreisch2020_PhRvD_101_3505, Hill2020_PhRvD_102_3507, DiValentino2020_NatAs_4_196, Muir2020_arXiv_2010_5924}.

The observed clustering properties of galaxies reflect the underlying cosmological matter field in which the galaxies evolve. The cosmological constraining power of galaxy clustering becomes especially stringent when spectroscopic redshift information is available, as redshift-space distortions (RSDs) can be used to probe the matter velocity field in addition to the density field. The primary constraint obtained from the analysis of redshift-space clustering is the cosmological parameter combination $f \sigma_8,$ where $f$ is the growth rate of structure obtained from linear perturbation theory, and $\sigma_8$ quantifies the normalisation of the matter power spectrum. The advent of large-scale galaxy surveys has opened up the road for such constraints, including from the the WiggleZ survey \citep{Blake2011_MNRAS_415_2876}, the \textit{Sloan Digital Sky Survey} \citep{Samushia2012_MNRAS_420_2102, Beutler2012_MNRAS_423_3430}, the \textit{Galaxy and Mass Assembly} survey \citep{Blake2013_MNRAS_436_3089}, the \textit{Baryon Oscillation Spectroscopic Survey} \citep[BOSS; see e.g.][]{Parejko2013_MNRAS_429_98,Alam2017_MNRAS_470_2617} and the \textit{Extended Baryon Oscillation Spectroscopic Survey} \citep[eBOSS; see e.g.][]{deMattia2020_MNRAS_tmp_3648, Bautista2021_MNRAS_500_736, Hou2021_MNRAS_500_1201}.

Theoretical predictions for the redshift-space clustering of galaxies are notoriously challenging. To leading order in perturbation theory, the RSD signal is determined by the Kaiser effect \citep{Sargent1977_ApJ_212_3, Kaiser1987_MNRAS_227_1}, i.e., the coherent infall of galaxies towards over-densities. However, even on very large scales, the precision of present-day galaxy surveys requires the RSD prediction to include a correction that accounts for the virial motion of satellites inside haloes \citep{Peacock1994_MNRAS_267_1020, Scoccimarro2004_PhRvD_70_3007}, the so-called \textit{Fingers of God} effect, a fundamentally nonlinear phenomenon that cannot be predicted from perturbation theory. 

It is well known that the constraining power of RSD analyses improves dramatically when including information from the quasi-to-nonlinear regime \citep[e.g.,][]{Zhai2019_ApJ_874_95}; the potential gains are so substantial that there now exists a substantial literature dedicated to the development of models that can extend the range of scales used in the analysis. This long-standing effort includes various extensions of perturbation theory \citep{Taruya2010_PhRvD_82_3522, Carlson2013_MNRAS_429_1674, Matsubara2014_PhRvD_90_3537, Wang2014_MNRAS_437_588}, the Gaussian streaming model \citep{Reid2011_MNRAS_417_1913}, the Zel'dovich streaming model \citep{White2014_MNRAS_439_3630}, Effective Field Theory \citep{Lewandowski2015_JCAP_05_019}, and approaches based on distribution functions \citep{Seljak2011_JCAP_11_039, Okumura2012_JCAP_11_014}. While this effort to extend linear theory has certainly improved the predictive power of contemporary analytical frameworks, the reliability of such efforts currently remains limited to scales larger than $\sim30 \, \hmpc$ \citep[see][for a review]{White2015_MNRAS_447_234}.

In a promising pilot study, \cite{Reid2014_MNRAS_444_476} highlighted the potential constraining power of extending the analysis of RSDs to smaller scales by performing an analysis on the BOSS CMASS sample. In the absence of reliable analytic models for redshift-space clustering on all scales, the authors used a single simulation with a re-scaling of the bulk velocity of haloes to constrain the growth rate. \cite{Reid2014_MNRAS_444_476} presented a $2.5\%$ constraint on $f \sigma_8$, a more than factor of two improvement over other studies using larger scales and on par with constraints on $S_8 = \sigma_8 \sqrt{\Omega_{\rm m} / 0.3}$, the equivalent cosmological quantity probed by large-scale structure probes involving gravitational lensing. Recently, \cite{Zhai2019_ApJ_874_95} made another significant step towards accurate constraints from non-linear scales by constructing a high-precision Gaussian Process (GP) emulator \citep[also see][]{Kwan2015_ApJ_810_35, Nishimichi2019_ApJ_884_29} to predict galaxy redshift-space clustering for BOSS CMASS. While \cite{Zhai2019_ApJ_874_95} argue that the re-scaling method of \cite{Reid2014_MNRAS_444_476} is subject to systematic errors and yields cosmological posteriors that are too narrow, they similarly show that including data from highly non-linear scales has the potential to significantly tighten constraints on the growth of structure.

On sufficiently large scales, the statistical relation between observed galaxies and the underlying density field can be described in terms of a perturbative bias expansion, with uncertainty in this relationship encoded by the coefficients of the expansion \citep[see][for a comprehensive review]{Desjacques2018_PhR_733_1}. Due to the non-linear physics of gravitational collapse, this formalism breaks down irreparably on sufficiently small scales, and one needs to directly model the relation between galaxies and the dark matter haloes that host them, colloquially known as the ``galaxy--halo connection''. There are two commonly used frameworks to model this relationship. In subhalo abundance matching \citep[SHAM; see e.g.][]{Conroy2006_ApJ_647_201, Lehmann2017_ApJ_834_37} one assumes a correspondence between all dark matter haloes, both field haloes and subhaloes accreted by other haloes, and galaxies. In contrast, in the halo occupation distribution \citep[HOD; see e.g.][]{Seljak2000_MNRAS_318_203, Berlind2002_ApJ_575_587, Zheng2007_ApJ_667_760, Hearin2016_MNRAS_460_2552, Sinha2018_MNRAS_478_1042} framework, or equivalently, the conditional luminosity function \citep[CLF; see e.g.][]{Yang2003_MNRAS_339_1057, vandenBosch2007_MNRAS_376_841} framework, one only models the relation between galaxies and field haloes but now allows field haloes to host multiple galaxies.

The goal of this paper is to perform the first consistent simulation-based cosmological RSD analysis of the full range of scales accessible with current large-scale structure data sets. Particularly, for this work, we analyse anisotropic redshift-space clustering in the scale range from $0.4 \, \hmpc$ to $63 \, \hmpc$. Our analysis is based on the Cosmological Evidence Modelling framework (CEM) introduced in \cite{Lange2019_MNRAS_490_1870}; in comparison to the widely used GP emulation technique \citep[e.g.,][]{Heitmann2010_ApJ_715_104}, the CEM provides improved flexibility to robustly quantify Bayesian posterior uncertainties with more realistically complex models of the galaxy--halo connection. We test our method using mock data sets created by populating dark matter haloes in numerical simulations with galaxies using the SHAM method, which we subsequently analyse using our HOD model applied to a different set of simulations. We demonstrate that our method is able to yield tight, unbiased cosmological constraints, which is a non-trivial result given that the SHAM model is substantially different from the HOD model. Having demonstrated that the HOD model is unbiased and sufficiently general, we next apply it to two carefully constructed galaxy samples from the BOSS LOWZ survey. After marginalising over uncertainties in the galaxy--halo connection, we obtain two independent measurements of $f \sigma_8$, each with an accuracy of $\sim 5\%$, a factor of two improvement over previous results. Finally, we quantify how our cosmological constraints depend on both small and large scales and explore predictions for gravitational lensing.

This paper is organized as follows. In section~\ref{sec:observations} we describe our observational data set from the BOSS LOWZ survey and the summary statistics we extract from it. We describe our modelling framework, including the HOD model and simulations used, in section \ref{sec:modelling}. Section \ref{sec:mock_tests} shows that our modelling framework is able to recover accurate cosmological constraints from mock measurements. Our cosmological constraints from applying the analysis framework to the observational data are described in section \ref{sec:results}. Finally, we discuss our results in section \ref{sec:discussion} and list our conclusions in \ref{sec:conclusion}.

Throughout this work, all observational measurements are made assuming a spatially flat $\Lambda$CDM cosmology with $\Omega_{\rm m}=0.307$.

\section{Observations}
\label{sec:observations}

\subsection{Sample selection}

We select galaxies from the BOSS LOWZ catalogue. Spectroscopic target selection for LOWZ requires that objects fulfil the following selection cuts:
\begin{eqnarray}
    r_{\rm cmod} &<& 13.5 + c_\parallel\, / 0.3 \label{eq:cut_c_parallel}\\
    |c_\perp| &<& 0.2 \label{eq:cut_c_perp}\\
    16 < &r_{\rm cmod}& < 19.6. \label{eq:m_cut}
\end{eqnarray}
In the above equations, $r_{\rm cmod}$ refers to the observed cmodel magnitudes and $c_\parallel$ and $c_\perp$ are colours defined as follows:
\begin{equation}
    c_\parallel = 0.7 (g_{\rm mod} - r_{\rm mod}) + 1.2 (r_{\rm mod} - i_{\rm mod} - 0.18)
\end{equation}
and
\begin{equation}
    c_\perp = r_{\rm mod} - i_{\rm mod} - (g_{\rm mod} - r_{\rm mod}) / 4.0 - 0.18\,.
\end{equation}
Note that these colours are measured using model magnitudes. We refer the reader to \cite{Eisenstein2001_AJ_122_2267} for a detailed motivation of these selection cuts and outline only the most salient points here. First, passively evolving galaxies form a locus in the $g - r$ versus $r - i$ plane. Equation (\ref{eq:cut_c_perp}) uses the position in the $g - r$ versus $r - i$ plane to select galaxies at $z \lesssim 0.45$. Additionally, equation (\ref{eq:cut_c_parallel}) results in a roughly redshift-independent cut on absolute magnitude. Finally, equation (\ref{eq:m_cut}) ensures that galaxies too faint to be targeted by the SDSS spectrograph are not selected. Ultimately, the above criteria for targets result in a luminous red galaxy (LRG) sample in the redshift range $z \sim 0.15 - 0.50$. We seek to select from this parent sample two approximately volume-limited samples of LRGs with median redshifts of $z \sim 0.25$ and $z \sim 0.40$, corresponding to the two Aemulus simulation outputs in the BOSS LOWZ redshift range.

We note that the above selection is based on apparent magnitudes. It is thus not expected that the target selection would be perfectly uniform in redshift. However, such a non-uniform sample selection violates the assumption of a redshift-independent HOD model that we will assume in the modelling in section~\ref{sec:results}. To create volume-limited samples of red galaxies we have to impose further cuts.

\begin{figure*}
    \centering
    \subfloat{\includegraphics{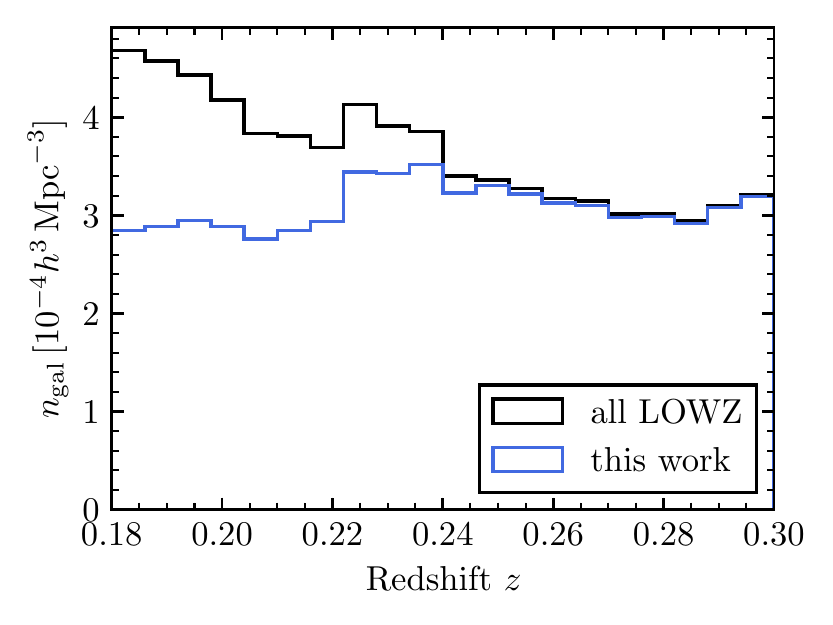}}
    \subfloat{\includegraphics{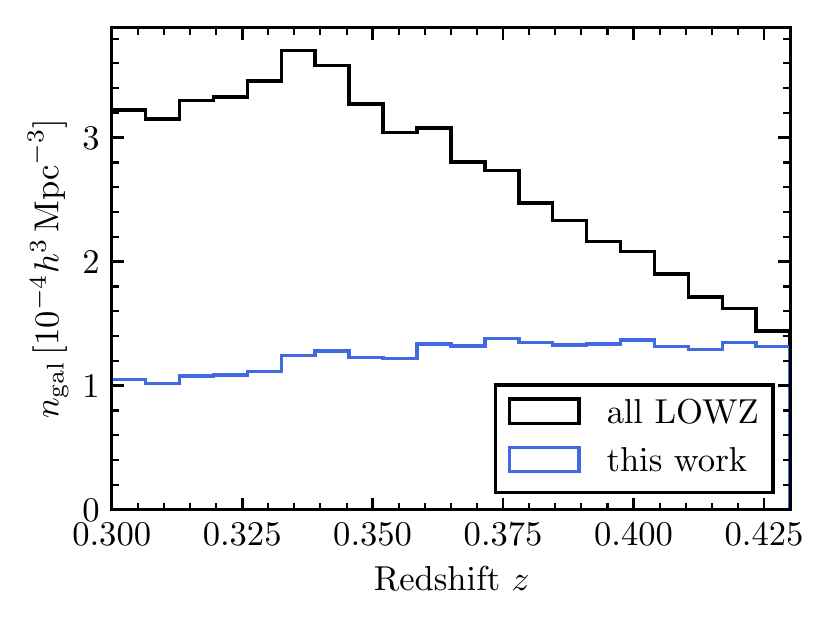}}
    \caption{The comoving galaxy number density of LOWZ galaxies in the redshift range $0.18 \leq z < 0.3$ (left) and $0.3 \leq z < 0.43$ (right). We show the distribution of the LOWZ parent sample (black) and our subset (blue).}
    \label{fig:z_dist}
\end{figure*}

We start by calculating $k$-corrected, absolute magnitudes at $z_0 = 0.25$ and $z_0 = 0.4$ for all BOSS LOWZ galaxies. The $k$-corrections are calculated from the model magnitudes using {\sc kcorrect} and, in the following, all absolute magnitudes are expressed in the AB magnitude system. Next, we select galaxies in the redshift ranges $(0.18, 0.30]$ and $(0.30, 0.43]$. Finally, we impose more stringent cuts on the absolute magnitudes ($M_{r, {\rm cmod}, z_0}$) and colours ($c_{\parallel, z_0}$ and $c_{\perp, z_0}$):
\begin{eqnarray}
    M_{r, {\rm cmod}, 0.25} &<& -25.874 + c_{\parallel, 0.25} / 0.3\\
    -0.216 < &c_{\perp, 0.25}& < 0.162\\
    M_{r, {\rm cmod}, 0.25} &<& -20.412
\end{eqnarray}
and
\begin{eqnarray}
    M_{r, {\rm cmod}, 0.40} &<& -26.899 + c_{\parallel, 0.40} / 0.3\\
    -0.154 < &c_{\perp, 0.40}& < 0.112\\
    M_{r, {\rm cmod}, 0.40} &<& -21.209
\end{eqnarray}
for the $0.18 < z \leq 0.30$ and $0.30 < z \leq 0.43$ samples, respectively. These cuts were calculated based on the average $k$-corrections as a function of redshift for the respective samples. In the absence of scatter in the $k$-corrections, all galaxies fulfilling these more stringent cuts also fulfil the general LOWZ target selection cuts.\footnote{We ignored that the LOWZ target selection requires $r_{\rm cmod} > 16$, thereby excluding objects that are too bright. However, there are only few physical objects for which this is true.}

Because these more stringent cuts are based on rest-frame properties, this should result in roughly volume-limited samples of red galaxies. In Figure~\ref{fig:z_dist}, we show the impact of our selection cuts on the comoving galaxy number density in both redshift ranges considered. Especially for the second sample, our selection cuts reduce the total number of galaxies in this range by a factor of $\sim 2$. The main reason for the reduction is the cut in absolute $r$-band magnitude. The original BOSS LOWZ selection effectively applied a very high cut on absolute magnitude at $z = 0.43$ and applying this across the redshift range reduces the number of galaxies at $z = 0.30$. At the same time, we see that our more stringent selection results in a roughly redshift-independent galaxy number density, as expected for a volume-limited sample without intrinsic number density evolution.

As described in \cite{Ross2017_MNRAS_464_1168}, there are slight differences in the photometric calibration of samples in the Northern Galactic Cap (NGC) and the Southern Galactic Cap (SGC). This results in slightly different target selections between the two hemispheres. Therefore, in this paper, we only select galaxies from the NGC. Overall, we have $79,018$ galaxies with a median redshift of $0.251$ for the first sample and $72,056$ galaxies with a median redshift of $0.377$ for the second sample. The corresponding comoving number densities are $(2.80 \pm 0.05) \times 10^{-4} \, h^{3} \, \mathrm{Mpc}^{-3}$ and $(1.18 \pm 0.02) \times 10^{-4} \, h^{3} \, \mathrm{Mpc}^{-3}$, respectively. The redshift ranges, $0.18 < z \leq 0.30$ and $0.30 < z \leq 0.43$, are narrow and correspond to time ranges of $1.2$ and $1.1 \, \mathrm{Gyr}$, respectively. Thus, we do not expect strong redshift evolution effects within each redshift bin.

\subsection{Clustering measurements}

\begin{figure*}
    \centering
    \subfloat{\includegraphics{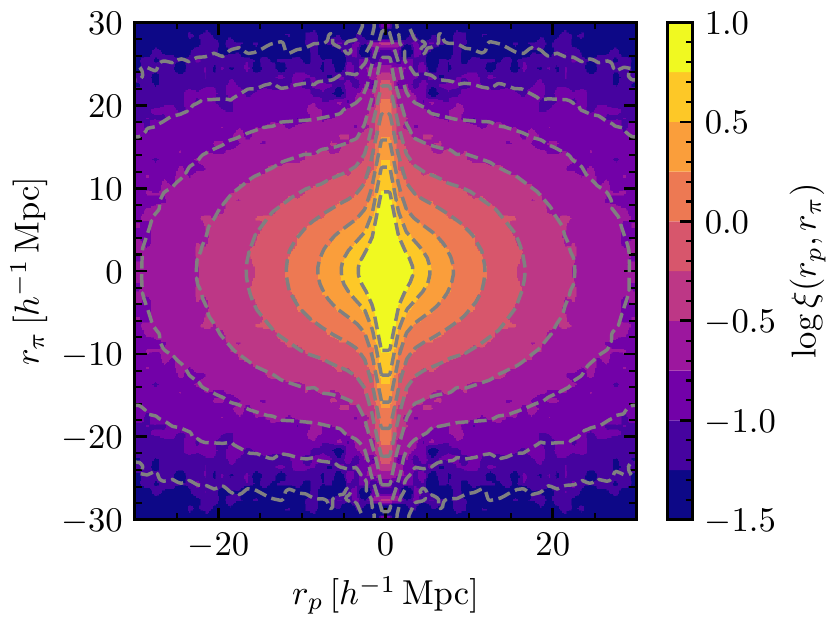}}
    \subfloat{\includegraphics{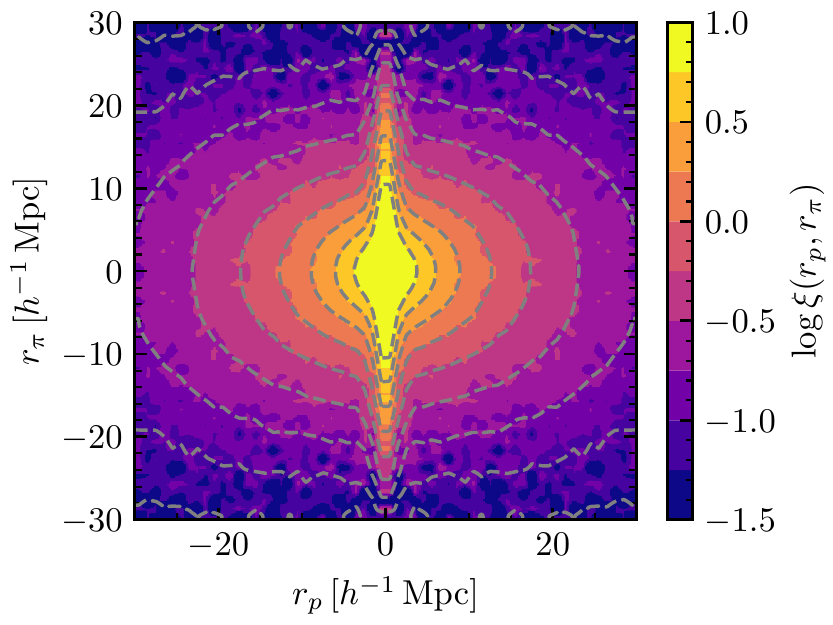}}
    \caption{The anisotropic two-point correlation function for the $0.18 \leq z < 0.3$ sample (left) and the $0.3 \leq z < 0.43$ sample (right). Colours indicate the measurements and dashed lines are predictions of the best-fit models discussed in section~\ref{sec:results} of where the colour transitions occur.}
    \label{fig:xi_rp_pi}
\end{figure*}

Following \cite{Guo2015_MNRAS_453_4368}, we measure the three-dimensional redshift-space two-point correlation function $\xi(s,\mu)$ through the Landy--Szalay estimator \citep{Landy1993_ApJ_412_64}, where $s$ is the redshift-space separation of the galaxy pairs and $\mu$ is the cosine of the angle between $s$ and the line of sight. We choose logarithmic $s$ bins with a width $\Delta\log s = 0.2$ from $0.1$ to $63.1 \, \hmpc$, and linear $\mu$ bins of width $\Delta\mu=0.05$ from $-1$ to 1. 

The multipole moments of order $\ell$ are then defined via
\begin{equation}
    \xi_\ell (s) = \frac{2\ell + 1}{2} \int\limits_{-1}^{1} L_\ell (\mu) \xi (s, \mu) d \mu \, ,
    \label{eq:multipoles}
\end{equation}
\citep{Hamilton1992_ApJ_385_5}. In the above equation $L_\ell$ represents the Legendre polynomial of order $\ell$. Note that contrary to \cite{Reid2014_MNRAS_444_476}, we integrate $L_\ell$ over the entire $\mu$-range. Measuring $\xi_\ell (s)$ on small scales $s$ is complicated by the presence of so-called fibre collisions, the fact that two spectroscopic fibres cannot be placed close to each other on a BOSS spectroscopic plate.

To accurately measure $\xi(s,\mu)$, we correct for the fibre collision effect of the BOSS LOWZ sample using the method of \cite{Guo2012_ApJ_756_127}. After being updated with the latest SDSS Data Release 16 \citep{Ahumada2020_ApJS_249_3}, the overall fraction of fibre collided galaxies is just $1.4\%$ for galaxies in the NGC of the BOSS LOWZ sample. The small-scale measurements can then be very accurately recovered, with minor fibre collision corrections.

The anisotropic two-point correlation functions are shown in Figure~\ref{fig:xi_rp_pi} for visualizations purposes. However, throughout this work, our modelling is performed using the multipole moments.

\subsection{Covariance matrix}
\label{subsec:covariance}

\begin{figure}
    \includegraphics{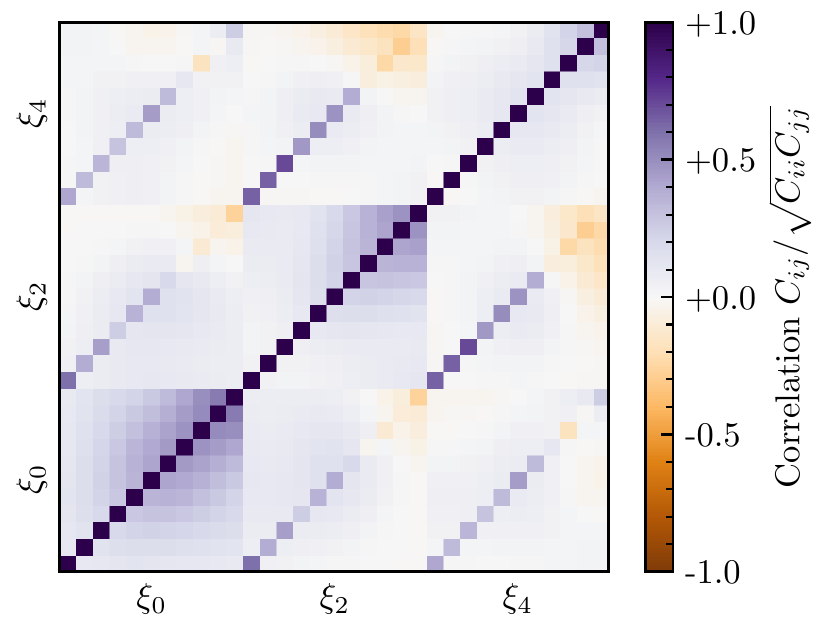}
    \caption{The assumed covariance matrix for the redshift-space clustering measurements. The scale $s$ increases from left to right and bottom to top. We see that multipole moments within the same $s$-bin are strongly correlated. In contrast, measurements at different scales $s$ only show significant correlation at larger scales. The covariance matrix here is for the sample at $z \sim 0.25$. The covariance matrix for the higher-redshift sample, $z \sim 0.4$, is qualitatively similar. See the text for details regarding the construction of the covariance matrix. We ignore cross-correlation between the clustering measurements and the galaxy number density.}
    \label{fig:covariance}
\end{figure}

Uncertainties on the measurements are derived from jackknife-resampling of $74$ roughly equal-area regions of the NGC. Note that this number of jackknife samples is comparable to the number of measurement points. Thus, without corrections, the inverse of the covariance matrix is expected to be a biased estimate of the precision matrix \citep{Hartlap2007_AA_464_399, Taylor2013_MNRAS_432_1928}. In principle, an unbiased estimate of the precision matrix can be obtained by applying the so-called Hartlap factor \citep{Hartlap2007_AA_464_399}. However, we opt for a more physically motivated correction to the covariance matrix estimator.

Let us look at the covariance matrix of two multipole moments $\xi_i$ and $\xi_k$. We expect the following statements to hold regardless of whether we study the covariance within the same multipole moment, $i = k$, or different moments, $i \neq k$. First, it is expected that the multipole moments of the same bin in $s$ are heavily correlated because, as shown in equation (\ref{eq:multipoles}), they are derived from the same $\xi$ measurements at fixed $s$ with different $\mu$ weights given by the Legendre polynomials. These correlations correspond to the diagonal elements of the covariance matrix and we choose not to modify those. Similarly, multipole moments should be less correlated if they do not belong to the same bin in $s$, i.e. off-diagonal elements, since they are derived from distinct measurements of $\xi(s, \mu)$. Additionally, it is expected that neighbouring off-diagonal elements of the covariance matrix that are close in scale show similar levels of correlation. Thus, we smooth the off-diagonal elements of the correlation matrix with a two-dimensional Gaussian kernel.\footnote{Before applying the smoothing, we replace each diagonal element with the average from the four neighbouring, off-diagonal elements.} The resulting correlation matrix is shown in Figure~\ref{fig:covariance} for the $z = 0.25$ sample. We choose a width of $2$ for the Gaussian kernel but the exact value does not significantly affect the goodness-of-fit of different theoretical models. Similarly, we find that this correction results in similar goodness-of-fit measures, i.e. $\chi^2$, as the Hartlap correction. We note that the approach employed here has been used in a very similar fashion in \cite{Mandelbaum2013_MNRAS_432_1544}, who also find that such a smoothing of the correlation matrix gives virtually the same results as applying the Hartlap factor to the noisy precision matrix.

When fitting models to the data we exclude scales below $400 \, \hkpc$. The signal-to-noise of the data on these small scales is poor, especially for the hexadecapole moments, and since the jackknife sub-samples have very few or no galaxy pairs at these separations, the corresponding estimates of the precision matrix are unreliable.

\section{Theoretical modelling}
\label{sec:modelling}

Our modelling is based on comparisons of the observational data described in the previous sections with predictions from simulations. In the following, we describe the simulations used and how we construct mock galaxy samples from them.

\subsection{Simulations}

Our modelling is based on the publicly available Aemulus simulations \citep{DeRose2019_ApJ_875_69}. Aemulus is a suite of $75$ cosmological, dark matter-only simulations with a volume of $\left( 1050 \, \hmpc \right)^3$ and $1400^3$ particles resulting in a particle mass resolution of $m_{\rm p} = 3.51 \times 10^{10} \sqrt{\Omega_{{\rm m}, 0} / 0.3} \, M_\odot / \mathrm{h}$. Our analysis is based on the first $40$ Aemulus simulations that all have different cosmologies. The cosmological parameters of all simulations are listed in the appendix and probe the $4 \sigma$ posterior ranges of the combined Baryon Acoustic Oscillation (BAO) plus CMB analysis by \cite{Anderson2014_MNRAS_441_24}.

The Aemulus simulation suite is specifically designed for the study of non-linear clustering in BOSS CMASS. The BOSS LOWZ and CMASS samples target luminous red galaxies of similar number densities and redshifts. Thus, the convergence analysis presented in \cite{DeRose2019_ApJ_875_69} should also be applicable to this work. Generally, particle resolution is the main limiting factor regarding the convergence of the Aemulus simulations. At fixed CMASS-like HOD, galaxy redshift-space correlation functions are converged to within $\sim 2.5\%$. This is better than the accuracy of our measurements except for $\xi_0$ over the range $1 \, \hmpc < s < 20 \, \hmpc$. Given that the analysis of \cite{DeRose2019_ApJ_875_69} is performed at fixed HOD, resolution effects could bias our galaxy--halo connection parameters. At the same time, it is not clear that small changes to the halo mass function or mass-concentration relation due to resolution effects would bias $f \sigma_8$ constraints significantly since they could easily be degenerate with galaxy--halo connection parameters. Indeed, in section \ref{sec:mock_tests} we perform mock tests on galaxy mock catalogues derived from the UNIT simulations which have $\sim 25$ times better mass resolution than the Aemulus simulations. Since we do not find biases in the inferred $f \sigma_8$ values from these mock tests, this serves as indication that the resolution of the Aemulus simulations is sufficient for our application.

\subsection{Halo catalogues}

Dark matter haloes in the simulations were identified with the {\sc ROCKSTAR} phase-space halo finder \citep{Behroozi2013_ApJ_762_109}. From the halo catalogue, we use field haloes with a mass of at least $100 \, m_{\rm p}$. We will use halo catalogues extracted at $z = 0.25$ and $z = 0.40$, as appropriate. Halo masses $M$ and radii $r_{\rm h}$ are defined using an over-density threshold of $\Delta = 200$ times the background density of the Universe at the redshift of the halo catalogues. Halo concentrations $c$ are derived from the maximum circular velocity $V_{\rm max}$ values provided in the catalogues assuming a spherically symmetric Navarro-Frenk-White (NFW) profile \citep{Navarro1996_ApJ_462_563}.

\subsection{Halo occupation}
\label{subsec:hod}

We use an HOD model to occupy the {\sc ROCKSTAR} halo catalogues with galaxies. Specifically, the average number of galaxies living in a halo of mass $M$ and $V_{\rm max}$ is assumed to be split into a central and a satellite galaxy component:
\begin{equation}
    \langle N_{\rm gal} | M, V_{\rm max} \rangle = \langle N_{\rm cen} | M, V_{\rm max} \rangle + \langle N_{\rm sat} | M, V_{\rm max} \rangle.
\end{equation}

The number of centrals at a given halo mass $M$ (averaged over $V_{\rm max}$) is parametrised by
\begin{equation}
    \langle N_{\rm cen} | M \rangle = \frac{f_\Gamma}{2} \left( 1 + \mathrm{erf} \left[ \frac{\log M - \log M_{\rm min}}{\sigma_{\log M}} \right] \right),
\end{equation}
where $f_\Gamma$, $\log M_{\rm min}$ and $\sigma_{\rm \log M}$ are free parameters. The above equation describes a function that smoothly transitions from $0$ to $f_\Gamma \leq 1$ with $\langle N_{\rm cen} | M \rangle = f_\Gamma / 2$ at $\log M_{\rm min}$. The rate of the transition is characterized by $\sigma_{\log M}$. Naively, one might expect that more massive haloes host brighter galaxies and thereby $\lim_{\log M \to \infty} = 1$ instead of $\lim_{\log M \to \infty} = f_\Gamma \leq 1$. However, the BOSS LOWZ target selection misses some bright galaxies due to colour and magnitude cuts \citep[][]{Leauthaud2016_MNRAS_457_4021}. Thus, we can expect the central occupation number of high-mass haloes to be below unity \citep[see e.g.][]{Leauthaud2016_MNRAS_457_4021,Hoshino2015_MNRAS_452_998,Guo2018_ApJ_858_30}.

The average number of satellites is given by
\begin{equation}
    \langle N_{\rm sat} | M \rangle = \left( \frac{M - M_0}{M_1} \right)^\alpha
\end{equation}
with $M_0$, $M_1$, and $\alpha$ being free parameters. The satellite number thus has a power-law dependence on halo mass with a lower cut-off characterized by $M_0$, particularly $\langle N_{\rm sat} | M \rangle = 0$ for $M \leq M_0$. Note that we do not implement a satellite galaxy analogue of $f_\Gamma$ that accounts for colour cuts in LOWZ centrals since such a free parameter would be degenerate with $M_1$.

In order to allow for galaxy assembly bias, we utilize the decorated HOD (dHOD) framework of \cite{Hearin2017_AJ_154_190} to parametrise the number of galaxies as a function of $V_{\rm max}$ at a fixed halo mass $M$. In this framework, the average number of galaxies is derived by modifying the average number at a fixed halo mass based on whether $V_{\rm max}$ is above or below the median $V_{\rm max}$  of all haloes at that mass,
\begin{equation}
    \langle N_{\rm gal} | M, V_{\rm max} \rangle = \langle N_{\rm gal} | M \rangle \pm \delta N_{\rm gal}.
    \label{eq:dHOD}
\end{equation}
In the above equation, $\delta N_{\rm gal}$ is added if $V_{\rm max}$ is larger than the median and subtracted otherwise. This parametrisation is not purely \textit{ad hoc} but reflects models of galaxy formation. For example, dark matter haloes that formed earlier, i.e. those with high $V_{\rm max}$, have less substructure and therefore likely fewer satellites \citep{Zentner2005_ApJ_624_505, Jiang2017_MNRAS_472_657}. In this case, we would expect $\delta N_{\rm sat} < 0$, at least for satellite galaxies. In the following, we will again separate contributions from central and satellite galaxies. For centrals, we use
\begin{equation}
    \delta N_{\rm cen} = A_{\rm cen} \left( 0.5 - \left| 0.5 - \langle N_{\rm cen} | M \rangle \right| \right),
\end{equation}
which fulfils the constraint $0 \leq \langle N_{\rm cen} | M, V_{\rm max} \rangle \leq 1$ for $-1 \leq A_{\rm cen} \leq 1$ as a free parameter. In contrast, for satellites we have
\begin{equation}
    \delta N_{\rm sat} = A_{\rm sat} \langle N_{\rm sat} | M \rangle,
\end{equation}
where $-1 \leq A_{\rm sat} \leq 1$ is another free parameter. With these average numbers specified, we assume that the number of centrals follows a Bernoulli distribution and satellites a Poisson distribution.

\subsection{Central galaxy phase-space coordinates}
\label{subsec:central_phase-space}

Central galaxies are assumed to coincide spatially with the halo centre defined as the average position of particles surrounding the dark matter halo phase-space density peak. Additionally, we assign a bulk velocity corresponding to the average velocity of all particles within $10\%$ of the halo radius $r_{\rm h} = R_{200 \rm b}$ around the density peak. We refer the reader to \cite{Behroozi2013_ApJ_762_109} for details regarding the phase-space positions of haloes. The definition of the bulk velocity used here is very similar to the $v_{\rm dens}$ definition in \cite{Reid2014_MNRAS_444_476}, which was shown to yield much better fits to available BOSS CMASS data than using the centre-of-mass velocity of each halo. Similarly, as shown in \cite{Ye2017_ApJ_841_45}, this definition traces the velocity of central galaxies in the Illustris hydrodynamical simulation much better than the centre-of-mass velocity.

Finally, following \cite{Reid2014_MNRAS_444_476} and \cite{Guo2015_MNRAS_446_578}, we add an additional Gaussian scatter to the line-of-sight component of the bulk velocity. The scatter has width $\sigma$,
\begin{equation}
    \sigma = \frac{\alpha_{\rm c} V_{\rm vir}}{\sqrt{3}} \, . 
\end{equation}
Here, $V_{\rm vir} = \sqrt{G M / r_{\rm h}}$ is the circular speed at the halo radius and $\alpha_{\rm c}$ is a free parameter. Our definition of central velocity bias is the same as in \cite{Reid2014_MNRAS_444_476}, where $\alpha_c$ is called $\gamma_{\rm cenv}$, and similar to \cite{Guo2015_MNRAS_446_578} with the difference being that the scatter is scaled by $V_{\rm vir}$ instead of the particle velocity dispersion. Physically, central velocity bias could originate from the central galaxy oscillating inside the dark matter halo or the dark matter halo not being fully relaxed \citep{vandenBosch2005_MNRAS_361_1203}. When defining central velocity bias with respect to the halo core as given in {\sc ROCKSTAR}, hydrodynamical simulations predict $\alpha_c \lesssim 0.1$ \citep{Ye2017_ApJ_841_45}.

\subsection{Satellite galaxy phase-space coordinates}
\label{subsec:satellite_phase-space}

We model satellite galaxy positions and velocities using parametrised analytical forms instead of placing them on resolved subhaloes or dark matter particles \citep[see e.g.][]{Reid2014_MNRAS_444_476, Guo2015_MNRAS_446_578, Guo2015_MNRAS_453_4368}. The number density of satellite galaxies within their host haloes is assumed to follow an NFW profile,
\begin{equation}
    n(r) \propto \frac{1}{r / r_s \left( 1 + r / r_s \right)^2} \, .
\end{equation}
In the above equation, $r_s$ is the scale radius and is expressed via $r_s = r_{\rm h} / c_{\rm sat}$, where $r_{\rm h}$ is the host halo radius. We allow the concentration parameter of satellites $c_{\rm sat}$ to be different than that of the dark matter $c_{\rm dm}$ via the free parameter $\eta$:
\begin{equation}
    c_{\rm sat} = \eta c_{\rm dm} \, .
\end{equation}
The average velocity of satellites is assumed to be the bulk velocity of the dark matter halo. Additionally, we add a stochastic Gaussian scatter to the line-of-sight velocity of each satellite. The amount of scatter depends on the distance $r$ from the halo centre and reflects satellite trajectories inside the dark matter halo. The width of the scatter is derived from solving the spherically symmetric Jeans equation without orbital anisotropy \citep{vandenBosch2004_MNRAS_352_1302} with an additional multiplicative factor $\alpha_{\rm s}$:
\begin{equation}
	\sigma^2 = \alpha_{\rm s}^2 \frac{G \eta^2 c^2 M}{r_{\rm h} g(c)} \left( \frac{r}{r_{\rm h}} \right) \left( 1 + \frac{\eta c r}{r_{\rm h}} \right)^2 \int\limits_{\eta c r / r_{\rm h}}^{\infty} \frac{g(y / \eta)\mathrm{d}y}{y^3 (1 + y)^2} \, ,
	\label{eq:Jeans}
\end{equation}
where $g(x) = \ln(1 + x) - x / (1 + x)$. The definition of satellite velocity bias is the same as in \cite{Reid2014_MNRAS_444_476}, \cite{Guo2015_MNRAS_446_578, Guo2015_MNRAS_453_4368} and \cite{Zhai2019_ApJ_874_95}, with $\alpha_s$ also being called $\gamma_{\rm IHV}$ or $\eta_{\rm vs}$, in the sense that it is a scaling of the satellite velocities. However, $\alpha_s$ is not the same as the ratio of the velocity dispersion of satellites and to that of dark matter particles. By solving the Jeans equation, this ratio is different from unity for $\eta \neq 1$ and $\alpha_{\rm s} = 1$. Instead, $\alpha_{\rm s} \neq 1$ describes a deviation from the prediction of the Jeans equation and can indicate that the kinematics of satellite galaxies is anisotropic, that satellite systems and/or dark matter haloes are not spherical, that satellite galaxies are not equilibrated within the halo potential, or any combination thereof. In general we thus have three free parameters, $\alpha_{\rm c}$, $\alpha_{\rm s}$ and $\eta$, to describe the phase-space coordinates of galaxies.  

\subsection{Clustering predictions}

We use the \texttt{s\_mu\_tpcf} and \texttt{tpcf\_multipole} functions from \texttt{halotools.mock\_observables} to predict redshift-space multipoles for mock galaxy populations. Particularly, we use the distant observer approximation while choosing one of the three axes of the simulation volume as the line of sight. In practice, we average results from projecting along each of the three simulation axes. The predictions do not take into account observational systematics like fibre collisions or survey boundary effects since we assume the measurements to be corrected for those. However, when making clustering predictions, we correct for the Alcock--Paczy\'nski (AP) effect \citep{Alcock1979_Natur_281_358}. The AP effect describes the phenomenon that inferred comoving coordinates of galaxies are derived from angular positions and redshift and thus depend on the reference cosmology assumed for this conversion. We take the AP effect into account by rescaling the phase-space coordinates of the mock galaxy population. Specifically, the line-of-sight coordinate is scaled via
\begin{equation}
    \pi_{\rm ref} = \pi_{\rm sim} \frac{E (z | \mathcal{C}_{\rm sim})}{E (z | \mathcal{C}_{\rm ref})}
    \label{eq:pi_stretch}
\end{equation}
and the perpendicular coordinates are scaled via
\begin{equation}
    r_{\rm p, ref} = r_{\rm p, sim} \frac{d_{\rm com} (z | \mathcal{C}_{\rm ref})}{d_{\rm com} (z | \mathcal{C}_{\rm sim})} \, .
    \label{eq:rp_stretch}
\end{equation}
In the above two equations, $\pi_{\rm ref}$ and $r_{\rm p, ref}$ are the coordinates that would be inferred using a reference cosmology $\mathcal{C}_{\rm ref}$ different from the simulation cosmology $\mathcal{C}_{\rm sim}$. The comoving distance $d_{\rm com}$ is expressed in $\hmpc$ and $E(z) = \sqrt{\Omega_{\rm m} (1 + z)^3 + (1 - \Omega_{\rm m}) (1 + z)^{3(1 + w_0)}}$. Consistent with section~\ref{sec:observations}, we use a spatially flat $\Lambda$CDM cosmology with $\Omega_{\rm m} = 0.307$ and $w_0 = -1$ as our reference cosmology.

We make use of {\sc TabCorr}\footnote{\url{https://github.com/johannesulf/TabCorr}} to speed up the predictions for mock observables. {\sc TabCorr} implements the method described in \cite{Zheng2016_MNRAS_458_4015} whereby correlation functions between haloes as a function of $M$ and $V_{\rm max}$ are tabulated. Afterwards, these correlation functions are convolved with $n_{\rm gal} (M, V_{\rm max})$ to get fast and accurate predictions for galaxy correlation functions. We refer the reader to \cite{Neistein2012_arXiv_1209_0463}, \cite{Reid2014_MNRAS_444_476}, \cite{Zheng2016_MNRAS_458_4015} and \cite{Lange2019_MNRAS_490_1870} for details.

We use $100$ bins in $\log M$ going from $100 \, m_p = 3.51 \times 10^{12} ( \Omega_m / 0.3 ) h^{-1} M_\odot$ to the maximum halo mass in the simulation and two bins for $V_{\rm max}$ at fixed $M$ corresponding to $V_{\rm max}$ values above and below the median. We have tested that the number of halo mass bins is sufficient given the precision of our observations. Halo correlation functions are tabulated for mock central--central, central--satellite and satellite--satellite pairs. When calculating these auto- and cross-correlation functions, each halo is assigned one central and a Poisson number of satellite galaxies with a mean of $M / 10^{13} \, h^{-1} \, M_\odot$, significantly larger than the number of satellites we expect to find per halo.

Finally, the correlation functions are tabulated for fixed values of the galaxy phase-space parameters $\alpha_{\rm c}$, $\alpha_{\rm s}$ and $\eta$. To implement smooth variations for these parameters we need to rely on interpolation. We first construct a sample of $200$ points in $\alpha_{\rm c}$, $\alpha_{\rm s}$ and $\log \eta$ filling the prior space described in Table~\ref{tab:priors}. The sample points are the centres of clusters obtained from $k$-means clustering of the prior space\footnote{\url{https://scikit-learn.org/stable/modules/clustering.html\#k-means}}. The space on which the $k$-means clustering algorithm is run extends slightly beyond the prior space to ensure that the sample points fully encompass it. We tabulate correlation functions for all $200$ sample points and obtain predictions for arbitrary values of $\alpha_{\rm c}$, $\alpha_{\rm s}$ and $\log \eta$ through linear, barycentric interpolation. We test the accuracy of this approach through leave-one-out cross-validation, i.e. we test the accuracy of the prediction for a point on the grid when excluding it from the interpolation process. We find that all observables are predicted to within at least $0.2 \sigma$ accuracy where $\sigma$ is the observational uncertainty. The actual uncertainty due to interpolation is likely smaller because random points of the posterior generally have smaller distances to the $200$ data points than a data point to its nearest neighbours. Thus, we can neglect errors related to interpolation because they would likely be negligible, i.e. $\sqrt{1 \, \sigma^2 + 0.2^2 \, \sigma^2} \approx 1.02 \, \sigma$.

\subsection{Cosmological evidence modelling}

Throughout this paper, we use the Cosmological Evidence Modelling (CEM) framework \citep{Lange2019_MNRAS_490_1870} to derive cosmological constraints. The posterior constraint on cosmological parameters $\mathcal{C}$ is obtained by marginalising the full posterior over the galaxy--halo connection parameters $\mathcal{G}$ described in sections \ref{subsec:hod}, \ref{subsec:central_phase-space} and \ref{subsec:satellite_phase-space}:
\begin{equation}
    P(\mathcal{C} | \mathbf{D}) \propto P(\mathcal{C}) \int \mathcal{L} (\mathbf{D} | \mathcal{C}, \mathcal{G}) P(\mathcal{G}) d\mathcal{G} = P(\mathcal{C}) \mathcal{Z} (\mathbf{D} | \mathcal{C})\, .
    \label{eq:evidence_cosmology}
\end{equation}
In the above equations, $\mathbf{D}$ denotes the observational data, $\mathcal{L}$ the likelihood and $P(\mathcal{C})$ and $P(\mathcal{G})$ the priors on cosmology and the galaxy--halo connection, respectively. Throughout this work, we assume flat priors, i.e. $P(\mathcal{C}) \equiv P(\mathcal{G}) \equiv 1$ with ranges described in Table~\ref{tab:priors} for $\mathcal{G}$ and the volume probed by the Aemulus simulation suite for $\mathcal{C}$. The integral $\mathcal{Z} (\mathbf{D} | \mathcal{C})$, which we call cosmological evidence but is also known as the Bayesian evidence or marginal likelihood, can be calculated for each simulation of the Aemulus simulation suite, as detailed in section \ref{subsec:evidence_calculation}. Thus, we have $40$ samples of $\mathcal{Z} (\mathbf{D} | \mathcal{C}) \propto P(\mathcal{C} | \mathbf{D})$ for different parameter combinations of $\mathcal{C}$. After developing and fitting a model for $P(\mathcal{C} | \mathbf{D})$ to the $40$ samples, we then have an estimate for the full cosmological posterior. Compared to the widely used emulation method \citep{Zhai2019_ApJ_874_95, Nishimichi2019_ApJ_884_29, Wibking2020_MNRAS_492_2872}, this shifts the problem from emulating predictions for multi-dimensional observables $\hat{\mathbf{D}}$ as a function of $\mathcal{C}$ and $\mathcal{G}$ to emulating a single number, the evidence, as a function of $\mathcal{C}$. The advantage of this approach is that, in principle, we can use arbitrarily complex models for $\mathcal{G}$. In contrast, the standard emulation approach will suffer from a decrease in the emulator accuracy with increasing complexity of $\mathcal{G}$. Additionally, the observables, i.e. $\xi_{0, 2, 4}$, can be complicated functions of $\mathcal{G}$ and $\mathcal{C}$. Thus, to emulate them one needs sophisticated, non-parametric approaches with large degrees of freedom like Gaussian Process emulation. However, for the cosmological evidence, as we show in section \ref{subsec:evidence_modelling}, simple parametric forms like multi-dimensional Gaussian functions are often sufficient. This has the potential of increasing the accuracy of the CEM method over the standard emulation technique.

\section{Tests on mock catalogues}
\label{sec:mock_tests}

Before analysing the data described in section \ref{sec:observations}, we conduct mock tests to ensure that our analysis method can yield unbiased cosmology results. The tests we seek to conduct go beyond the investigations carried out by \cite{Zhai2019_ApJ_874_95} and \cite{Lange2019_MNRAS_490_1870}. These studies showed that unbiased cosmology results can be obtained when mock observations are created from the same model that is used to analyse the data. This result is non-trivial given the difficulty in predicting mock observables for arbitrary cosmology and HOD parameters. However, these findings do not show that unbiased cosmology results can be obtained for arbitrary galaxy populations. The main reason is that realistic galaxy populations will violate several of the assumptions made in sections~\ref{subsec:hod},~\ref{subsec:central_phase-space}~and~\ref{subsec:satellite_phase-space} \citep[see, e.g.][]{Hadzhiyska2020_MNRAS_493_5506, BeltzMohrmann2020_MNRAS_491_5771}. What remains unclear is to what extent these differences can bias our findings regarding cosmology here.

\subsection{SHAM galaxy model}

To explore to what extent complex galaxy populations can bias our cosmological inference, we utilize mock catalogues based on subhalo abundance matching (SHAM) \citep[see][for a review]{Wechsler2018_ARAA_56_435}. Specifically, we use the SHAM model first introduced by \cite{Lehmann2017_ApJ_834_37}. Briefly, this model populates all dark matter (sub-)haloes above a threshold in $V_\alpha$ with galaxies. Here, $V_\alpha$ is defined via the virial velocity $V_{\rm vir}$ and the maximum circular velocity $V_{\rm max}$ of a dark matter (sub-)halo,
\begin{equation}
    V_\alpha = V_{\rm vir} \left( \frac{V_{\rm max}}{V_{\rm vir}} \right)^\alpha \, .
\end{equation}
The free variable $\alpha$ determines the degree to which the matching depends on the concentration of the dark matter halo. For $\alpha \neq 0$, $V_\alpha$ depends on both halo mass and concentration. Ultimately, the average occupation and moments constructed using this SHAM model will deviate from the parametrised forms assumed for our HOD model, as described in section ~\ref{subsec:hod}. Furthermore, satellite galaxies in this model follow the dynamics of resolved subhaloes in simulations. Thus, we expect them to violate some of the assumptions made in section~\ref{subsec:satellite_phase-space} such as spherical symmetry or orbital isotropy. However, since the SHAM is performed on dark matter-only simulations, the mock catalogues do not include the effects of baryonic feedback on the matter distribution and dynamics. We leave the study of these effects on RSD measurements for future work.

\subsection{Mock measurements}

Our mock measurements are based on applying the SHAM model described above to the UNIT simulations \citep{Chuang2019_MNRAS_487_48}. Specifically, we use four simulations with a fixed power-spectrum normalization but complementary phases. Each of these simulations has $4096^3$ dark matter particles in a volume of $(1 \, h^{-1} \, \mathrm{Gpc})^3$. We study the simulation snapshots at redshift $z = 0.39$. The cosmological parameters of these simulations follow the \citet{PlanckCollaboration2016_AA_594_13} CMB analysis, i.e. $\Omega_m = 0.3089$, $h = H_0 / 100 = 0.6774$, $n_s = 0.9667$ and $\sigma_8 = 0.8147$. Thus, the $f \sigma_8 (z = 0.39)$ value we seek to recover is $0.476$.

The mock galaxy sample is constructed by setting the SHAM model parameter $\alpha$ to $0.73$ as motivated by the findings of \cite{Lehmann2017_ApJ_834_37}. We have also tested our results with $\alpha = 0$ and $\alpha = 1.5$, finding nearly identical cosmological constraints, as described below. Instead of applying a hard cut-off in $V_\alpha$ for each (sub)halo, we use a smooth transition. For field haloes, the probability to host a galaxy is $50\%$ at $V_\alpha \approx 700 \, \mathrm{km} \, \mathrm{s}^{-1}$ and increases roughly linear with a rate of $\sim 0.1 \, \% \, \mathrm{km}^{-1} \, \mathrm{s}$. This simulates scatter between halo and galaxy properties. For subhaloes, we require a slightly higher $V_\alpha \approx 760 \, \mathrm{km} \, \mathrm{s}^{-1}$ for a $50\%$ chance to host a satellite. We choose these slightly different $V_\alpha$ values for centrals and satellites in order for the mock clustering measurements to roughly match the observations of the $0.3 < z \leq 0.43$ sample. Additionally, they are chosen such that the mock catalogues reproduce the number density in observations of the same sample.

\begin{figure}
    \includegraphics{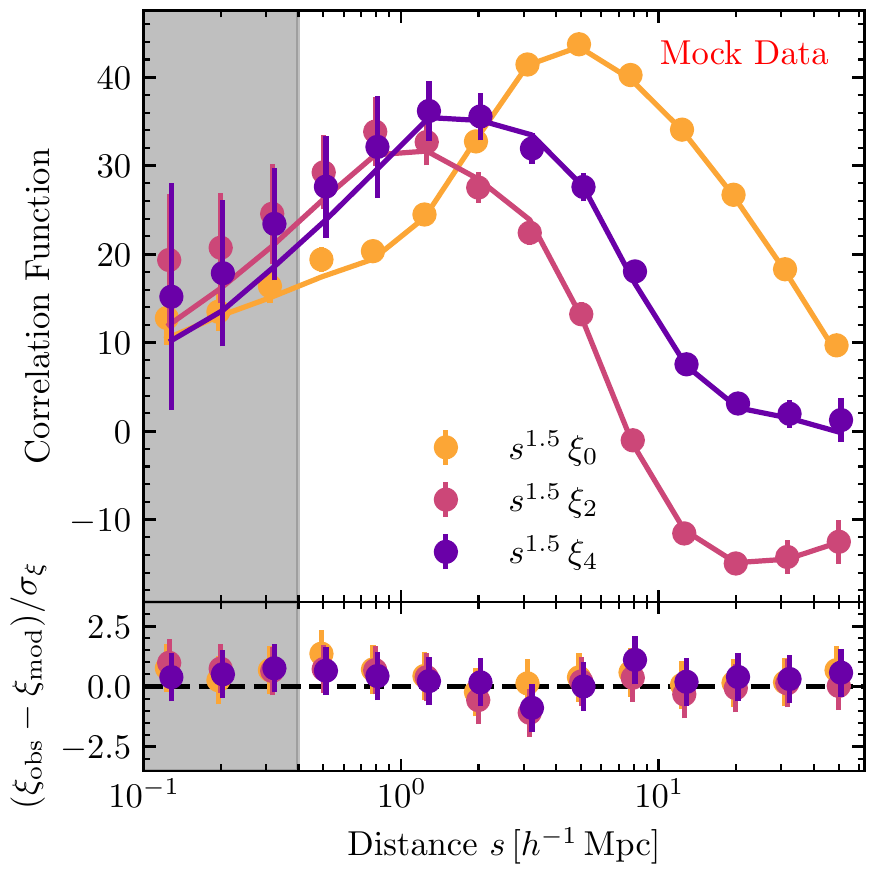}
    \caption{Mock measurements of the multipole moments of redshift-space correlation function. The measurements are based on applying a SHAM model to the UNIT simulations. Since the input cosmology is known, we use these measurements to test our analysis pipeline. The best-fit HOD model using the Aemulus simulations is shown by the solid line. All quantities are expressed in units of $\hmpc$.}
    \label{fig:mock_measurements}
\end{figure}

Mock measurements are obtained for the same statistics as for the observations. We calculate the redshift-space clustering by projecting each of the four simulations along each of the three spatial axes, $x$, $y$ and $z$. As our final mock data vector, we take the mean of the $12$ mock measurements. As the covariance matrix, we use the same covariance matrix as for the $0.3 < z \leq 0.43$ BOSS sample. The resulting mock measurements and uncertainties are shown in Figure~\ref{fig:mock_measurements}. In the same figure, we also show the best fit to the mock data using the HOD model. Specifically, the fit shown is the best fit obtained over all $40$ Aemulus $z = 0.40$ simulations outputs after marginalising over all HOD parameters listed in Table~\ref{tab:priors}. Because the covariance matrix reflects our observational volume of $\sim 0.6 \, (\mathrm{Gpc} / h)^3$ instead of the $4 \, (\mathrm{Gpc} / h)^3$ volume of the mock data, the best-fit $\chi^2 = 8.2$ is significantly smaller than the number of measurement points.

\subsection{Evidence calculation}
\label{subsec:evidence_calculation}

\begin{table}
    \centering
    \begin{tabular}{c|c|c}
    Parameter & Minimum & Maximum \\
    \hline
    $\log M_{\rm min}$ & 12.5 & 14.0 \\
    $\log M_0$ & 12.0 & 15.0 \\
    $\log M_1$ & 13.5 & 15.0 \\
    $\sigma_{\log M}$ & 0.1 & 1.0 \\
    $\alpha$ & 0.5 & 2.0 \\
    $f_\Gamma$ & 0.5 & 1.0 \\
    $A_{\rm cen}$ & -1.0 & 1.0 \\
    $A_{\rm sat}$ & -1.0 & 1.0 \\
    $\log \eta$ & $-\log 3$ & $+\log 3$ \\
    $\alpha_{\rm c}$ & 0.0 & 0.4 \\
    $\alpha_{\rm s}$ & 0.8 & 1.2 \\
    \end{tabular}
    \caption{Prior limits for all galaxy--halo connection parameters assumed when calculating the evidence. We always assume flat priors within the limits stated above.}
    \label{tab:priors}
\end{table}

We use the CEM method to extract cosmological information from the mock data using the Aemulus simulation suite. As a first step, we define the likelihood via
\begin{equation}
    \ln \mathcal{L} (\mathcal{C}, \mathcal{G}) = \frac{(n_{\rm gal} - \hat{n}_{\rm gal})^2}{2 \sigma_{n_{\rm gal}}^2} + \frac{1}{2} (\xi - \hat{\xi})^{\rm T} \Sigma^{-1 } (\xi - \hat{\xi}) \, .
    \label{eq:likelihood}
\end{equation}
In the above equation, $\xi$ contains the monopole, quadrupole and hexadecapole measurements, i.e. $\xi = [\xi_0, \xi_2, \xi_4]$. $\Sigma$ is the covariance matrix of the LOWZ measurements and $\sigma_{n_{\rm gal}}$ is the observational error on the number density. Finally, $\mathcal{G}$ and $\mathcal{C}$ denote the parameters describing the galaxy--halo connection model and cosmology, respectively. Finally, to calculate the integral in equation (\ref{eq:evidence_cosmology}) we use {\sc MultiNest} using Importance Nested Sampling \citep{Feroz2010_arXiv_1001_0719} with $10000$ live points and a target efficiency of $4\%$.

\subsection{Evidence modelling}
\label{subsec:evidence_modelling}

\begin{figure}
    \includegraphics{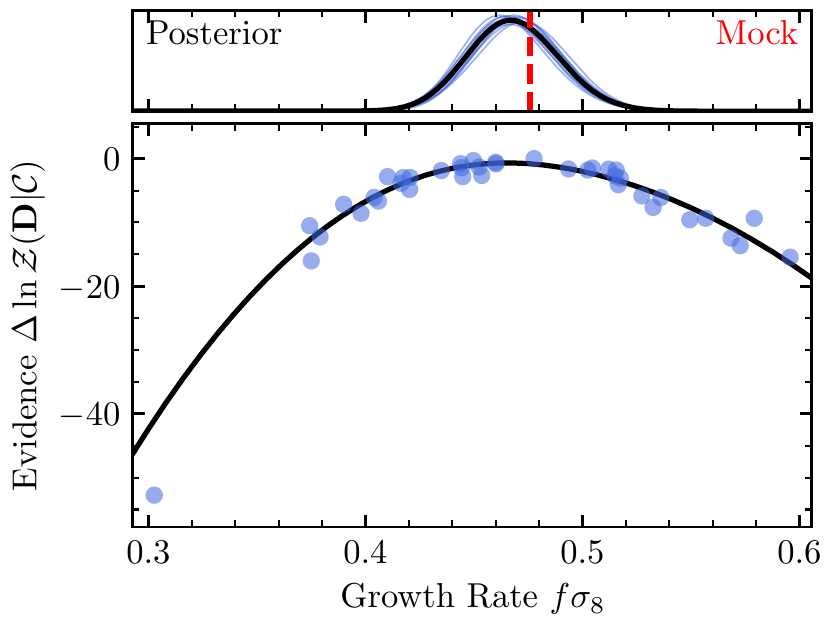}
    \caption{CEM analysis of mock data. The lower panels shows the cosmological evidence $\mathcal{Z} (\mathbf{D} | \mathcal{C})$ as a function of $f \sigma_8$ for the 40 regular simulations of the Aemulus simulation suite. The evidence is calculated with respect to the mock observations described in section~\ref{sec:mock_tests}. The line shows the best-fit skew normal to the evidence values. The upper panel shows as thin lines random draws from the posterior of skew normal distributions fitted to the evidence values. Finally, the solid black line in the upper panel is the superposition of all skew normal distributions of the posterior serving as the effective posterior constraint on $f \sigma_8$. This posterior constraint on $f \sigma_8$ compares favourably to the input value (red dashed) in the mocks that we seek to recover.}
    \label{fig:log_ev_fs8_challenge}
\end{figure}

In the lower panel of Figure~\ref{fig:log_ev_fs8_challenge}, we show a scatter plot of the evidence $\mathcal{Z} (\mathbf{D} | \mathcal{C})$ as a function of $f \sigma_8 (z=0.40)$ for each of the 40 regular simulations of the Aemulus simulation suite. From linear theory, we expect observations of the redshift-space correlation function on large, linear scales to primarily constrain $f \sigma_8$. Thus, it is natural to expect the same to hold true on non-linear scales. And indeed, Figure~\ref{fig:log_ev_fs8_challenge} shows that the evidence is a strong function of $f \sigma_8$. As discussed in \cite{Lange2019_MNRAS_490_1870}, because the cosmological evidence is directly proportional to the posterior on cosmology, once we have a reliable model for $\mathcal{Z} (\mathbf{D} | \mathcal{C})$, we also have obtained posterior constraints on cosmology.

\subsubsection{One-dimensional models}

For the moment, we follow \cite{Lange2019_MNRAS_490_1870} and assume that $\mathcal{Z} (\mathbf{D} | \mathcal{C})$ is a function of $f \sigma_8$ only. We would naively expect the posterior constraint on $f \sigma_8$ to be similar to a normal distribution. A skew normal distribution is a natural extension by allowing non-zero skewness. This additional freedom is found in \cite{Lange2019_MNRAS_490_1870} to be necessary to fit $\mathcal{Z} (\mathbf{D} | \mathcal{C})$ for other mock RSD data. Thus, this functional form is primarily empirically motivated. The skew normal distribution is parametrised by
\begin{equation}
    \begin{split}
        \mathcal{Z} (\mathbf{D} | \mathcal{C}&) \approx \hat{\mathcal{Z}} (\mathbf{D} | f \sigma_8)\\\propto& \left[ 1 + {\rm erf} \left( \frac{\alpha (f \sigma_8 - \mu)}{\sqrt{2} \sigma} \right) \right] \exp \left[ - \frac{(f \sigma_8 - \mu)^2}{2 \sigma^2} \right] \, .
        \label{eq:evidence_fsigma8}
    \end{split}
\end{equation}
Here, $\mu$, $\sigma$ and $\alpha$ are free parameters to be obtained by fitting the distribution shown in the lower panel of Figure~\ref{fig:log_ev_fs8_challenge}. We note that some simulations may not be able to yield reasonable fits to the data, thus resulting in very low evidence values. Since the cosmological parameters of such simulations would be ruled out anyway, we do not want them to influence the fit to the parametric evidence form $\hat{\mathcal{Z}}$ since this form is purely empirically motivated and might not hold over large ranges in $f \sigma_8$. Thus, we remove all simulations for which the best-fit predicted evidence $\ln \hat{\mathcal{Z}}$ value is $25$ below the maximum best-fit evidence. This roughly corresponds to simulations that are more than $7 \sigma$ away from the maximum posterior. In the present case, this only removes one simulation for which $f \sigma_8 \sim 0.3$.

From Figure~\ref{fig:log_ev_fs8_challenge}, it is also apparent that any reasonable parametric form for $\hat{\mathcal{Z}}$ will not perfectly fit the evidence values of all simulations. Some amount of scatter between the predicted and the measured $\mathcal{Z} (\mathbf{D} | f \sigma_8)$ is expected and can be due to a variety of reasons. For example, part of the scatter could be due to our approximation that the evidence values only depend on $f \sigma_8$. However, even if the evidence value was purely a function of $f \sigma_8$, a non-zero scatter would be expected due to cosmic variance. Each simulation probes a finite cosmological volume. Thus, the model predictions of each simulation will be affected by the random realizations of the phases and amplitudes of the perturbation modes in the initial conditions of each simulation. Furthermore, sampling noise when probing the phase-space properties of galaxies (see sections ~\ref{subsec:central_phase-space} and ~\ref{subsec:satellite_phase-space}) as well as errors when interpolating between $\alpha_s$, $\alpha_c$ and $\eta$ values could further add to the scatter. For the moment, we assume that the main source of the scatter in $\mathcal{Z} (\mathbf{D} | \mathcal{C})$ is truly random, i.e. cosmic variance, of the simulations themselves. We present further evidence for this assumption in the appendix. If the assumption holds, \cite{Lange2019_MNRAS_490_1870} advocated for the following functional form for the evidence scatter,
\begin{equation}
    \Delta \ln \mathcal{Z} \approx \sqrt{\frac{r_{\rm sim}^2 N_{\rm data}}{2} + r_{\rm sim} \chi_{\rm min}^2}.
\end{equation}
Here, $N_{\rm data}$ is the number of data points, $\chi_{\rm min}^2$ the best-fit $\chi^2$ for each simulation and $r_{\rm sim}$ another free parameter. In essence, the above functional form allows for arbitrary scatter values that increase with increasing $\chi^2$ of each simulation.

We choose flat priors for $\mu$, $\sigma$, $2 / \pi \arctan \alpha$ and $r_{\rm sim}$ with ranges $[0.35, 0.60]$, $[0.0, 0.1]$, $[-1, 1]$ and $[0, 1]$, respectively. When fitting $\mathcal{Z}(\mathbf{D} | f \sigma_8)$, we obtain a posterior sample of skew normal distributions that could fit the evidence values in Figure~\ref{fig:log_ev_fs8_challenge}. In order to account for uncertainties in the fit, we choose as the final cosmological posterior, the average of the skew normal distributions in our posterior sample. The upper panel of Figure~\ref{fig:log_ev_fs8_challenge} shows as thin blue lines random draws from the posterior and as a thick black line the total posterior. Overall, we infer $f \sigma_8 = 0.468 \pm 0.021$, comparing favourably to the input value $f \sigma_8 = 0.476$.

\subsubsection{Two-dimensional models}

\begin{figure}
    \includegraphics{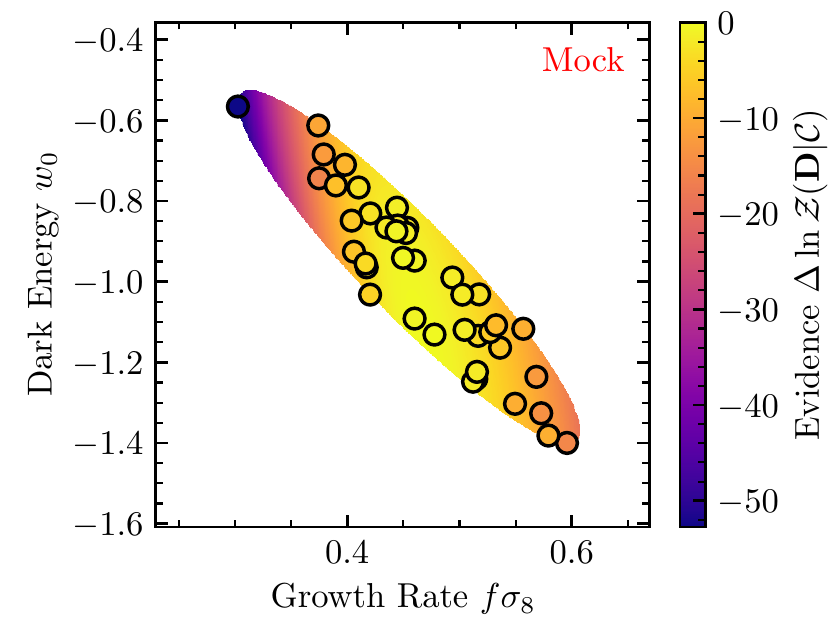}
    \caption{Cosmological evidence as a function of $f \sigma_8$ and the dark energy equation-of-state parameter $w_0$ for the mock data. Each point corresponds to one of the $40$ Aemulus simulations. The background colour in the ellipse is the best-fit model to the points.}
    \label{fig:log_ev_fs8_w0_challenge}
\end{figure}

In the one-dimensional model for $\mathcal{Z} (\mathbf{D} | \mathcal{C})$, $\mathcal{Z} (\mathbf{D} | \mathcal{C}) \approx \hat{\mathcal{Z}} (\mathbf{D} | f \sigma_8)$, we have implicitly assumed that any deviation from the best-fit model comes purely from random errors associated with cosmic variance. However, it is possible that scatter around the best-fit $\mathcal{Z}(\mathbf{D} | f \sigma_8)$ comes from unmodelled cosmological parameter dependencies. For example, there could be a degeneracy between $f \sigma_8$ and other cosmological parameters. Neglecting such a degeneracy could, in principle, result in artificially narrow posterior constraints in $f \sigma_8$. Here, we test for such a degeneracy by modelling the evidence with two-dimensional models, i.e. $\mathcal{Z}(\mathbf{D} | \mathcal{C}) = \mathcal{Z}(\mathbf{D} | f \sigma_8, \gamma)$ with $\gamma$ being a suitable cosmological parameter.

As the functional form for $\mathcal{Z}(\mathbf{D} | f \sigma_8, \gamma)$ we choose a generalization of a two-dimensional Gaussian distribution that is skew normal along one of the two axes. First, we centre and normalize the values of $f \sigma_8$ and $\gamma$ of the simulations, i.e. $\hat{\gamma} = (\gamma - \langle \gamma \rangle) / \sigma_\gamma$. We then obtain new coordinates $c_1$ and $c_2$ from a rotation of the normalized cosmological parameters, i.e. $[c_1, c_2]^T = \mathbf{R} [\widehat{f \sigma_8}, \hat{\gamma}]^T$. The rotation angle $\phi$ of the rotation matrix $\mathbf{R}$ is a free parameter. Finally, the total evidence is the product of a normal and a skew normal,
\begin{equation}
    \mathcal{Z} (\mathbf{D} | \mathcal{C}) = \mathrm{Skew} (c_1 | \mu_1, \sigma_1^2, \alpha) \times \mathrm{Normal} (c_2 | \mu_2, \sigma_2^2)
\end{equation}
For $\alpha = 0$, the above equation describes the probability density function of an ordinary two-dimensional Gaussian distribution. Similarly, for $\phi = 0$ and $\sigma_2^2 \to \infty$, this two-dimensional functional form reduces to the one-dimensional model for $\mathcal{Z} (\mathbf{D} | \mathcal{C})$. We choose flat priors for $\phi$, $\mu$ and $\sigma$ with ranges of $[-\pi/4, +\pi/4]$, $[-3, +3]$ and $[0.05, 3.0]$, respectively. The priors on $\alpha$ and $f_{\rm sim}$ remain unchanged.

In Figure~\ref{fig:log_ev_fs8_w0_challenge}, we show a two-dimensional analysis in which the second parameter $\gamma$ is the dark energy equation-of-state parameter $w_0$. As previously, each point corresponds to one simulation of the Aemulus simulation suite. The colour indicates the evidence value associated with each simulation. The background colour now shows the best-fit two-dimensional model. We can now, in principle, infer constraints on $f \sigma_8$ by marginalising over the $w_0$ direction and proceeding in the same way as for the one-dimensional evidence model. However, doing so naively would result in very weak constraints on $f \sigma_8$. The reason is that the fit prefers evidence models where the evidence depends on only one cosmological dimension, $f \sigma_8 + \epsilon w_0$. While the fit prefers $\epsilon = 0$, i.e. the evidence only depends on $f \sigma_8$, $\epsilon \neq 0$ cannot be ruled out. Thus, if no prior on $w_0$ was applied, constraints on $f \sigma_8$ would be weak because $w_0$ and $f \sigma_8$ might be slightly degenerate and no constraint on $w_0$ is found. However, because $\epsilon \sim 0$, even modest priors on $w_0$ are enough to prevent this. For example, it makes sense to explicitly impose a prior related to the cosmological parameter space probed by the simulations. We do so by constructing the minimum bounding ellipse around simulation parameters in the $f \sigma_8 - w_0$ plane, as shown in Figure~\ref{fig:log_ev_fs8_w0_challenge}. After imposing this strict prior on $f \sigma_8$ and $w_0$, we infer a marginalised constraint of $f \sigma_8 = 0.467 \pm 0.021$, in very good agreement with the one-dimensional result.

\subsection{Accuracy of cosmological constraints}

\begin{figure}
    \includegraphics{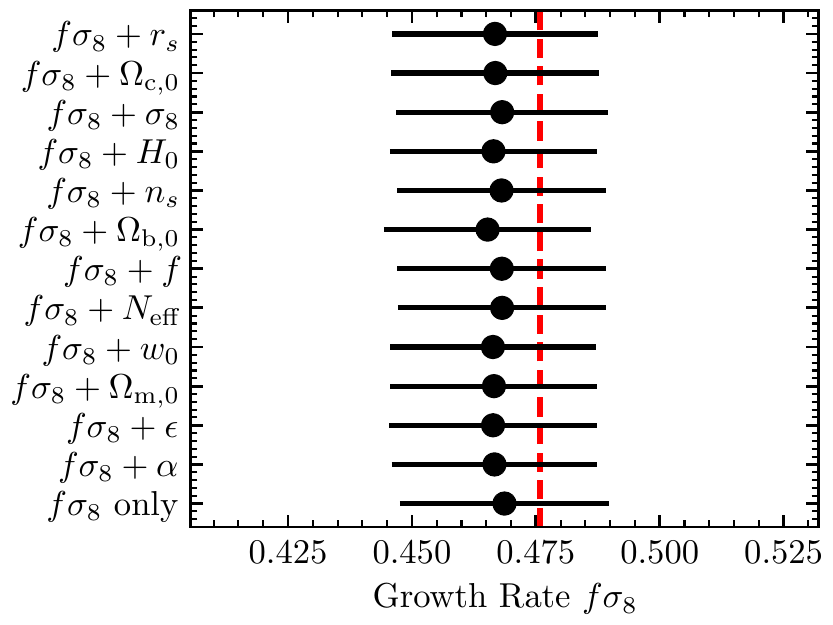}
    \caption{Marginalised constraints on $f \sigma_8$ for the mock as a function of the model for the evidence. The lowest error bar shows the $f \sigma_8$ constraint if the evidence is assumed to only depend on $f \sigma_8$. All other error bars show results if the evidence is assumed to depend on $f \sigma_8$ and an additional cosmological parameter. The input value from mock data set is shown by the red dashed line.}
    \label{fig:growth_rate_challenge}
\end{figure}

Figure~\ref{fig:growth_rate_challenge} shows how the marginalised constraints on $f \sigma_8$ depend on modelling choices. Specifically, the lowest error bar shows the result for the one-dimensional model and all other error bars results of two-dimensional models. In this figure, $\epsilon$ and $\alpha$ refer to the distortion parameters of the AP effect \citep{Padmanabhan2008_PhRvD_77_3540}. Most importantly, we see that all analysis methods result in very similar marginalised $f \sigma_8$ constraints. All shifts in the mean of the posterior are negligible and the uncertainty on $f \sigma_8$ is virtually unaffected. This seems to suggest that our modelling of the cosmological evidence is robust, even when assuming it only depends on $f \sigma_8$.

For simplicity, we choose the result from the one-dimensional modelling as the final result on $f \sigma_8$. Note that our mock observations are constructed using a volume of $4 (h^{-1} \, \mathrm{Gpc})^3$ whereas the covariance matrix corresponds to a volume of $\sim 0.6 (h^{-1} \, \mathrm{Gpc})^3$. Furthermore, for the SHAM mock catalogues, we project onto the three simulation axes. Since most of our constraining power comes from small scales, as shown later, this effectively triples the volume \citep{Smith2021_MNRAS_500_259}. Thus, we should expect to be able to recover the input $f \sigma_8 = 0.476$ to within roughly $\pm \, 0.021 / \sqrt{3 \times 4 / 0.6} = 0.005$. At the same time, there is additional uncertainty in the best-fit evidence model $\hat{\mathcal{Z}}(f \sigma_8)$. For example, the maximum evidence value is inferred to occur at $f \sigma_8 = 0.467 \pm 0.004$. Ultimately, the difference between the maximum $f \sigma_8$ evidence and the input is $0.009$, too small to confidently claim a detection of a bias given uncertainties in the mock observables and evidence model. Larger simulations, both for the mocks and for the modelling, would be needed to robustly detect a possible bias. Overall, if our recovered values for $f \sigma_8$ are biased, the bias is likely lower than statistical uncertainties for $f \sigma_8$ from current observations.

We have repeated the above exercise with $\alpha = 0$ and $\alpha = 1.5$ when constructing the SHAM mock. Changing $\alpha$ impacts the galaxy assembly bias strength and the satellite fraction of the mock sample \citep{Lehmann2017_ApJ_834_37}. Consequently, we find that the multipoles change significantly. For example, the quadrupole $\xi_2$ on small scales changes by $\sim 20\%$. Encouragingly, despite these significant changes in the observables, we find nearly identical $f \sigma_8$ constraints, $f \sigma_8 = 0.469 \pm 0.021$ and $f \sigma_8 = 0.467 \pm 0.021$. The small shift in $f \sigma_8$ makes sense given that the three mocks are derived from the same halo catalogue. Thus, the mock observations constructed using different values for $\alpha$ are likely highly correlated. In addition to changing $\alpha$, we performed a similar mock test for an analogue to the $0.18 < z \leq 0.30$ sample that we derived from the $z = 0.25$ UNIT simulation outputs and analyse with the $z = 0.25$ Aemulus outputs. Using $\alpha = 0.73$ for this low-redshift mock we find $f \sigma_8 = 0.476 \pm 0.022$ for an input values of $0.469$. As before, the recovery of $f \sigma_8$ is satisfactory given the volume of the mock observations and uncertainties in the evidence model. Overall, our tests with SHAM mocks do not indicate any significant biases in the recovery of $f \sigma_8$.

\subsection{Impact of redshift evolution}

The observations described in section~\ref{sec:observations} cover a range of redshifts while the Aemulus simulation snapshots we use to model them are coming from a single redshift. To mitigate the impact of potential biases we construct an observational galaxy sample that is roughly volume-limited and use simulation snapshots that correspond to approximately the mean redshift of galaxies in the sample. Nonetheless, we want to test here whether having observations taken at slightly different redshifts than the simulation output can bias our inferences. If our modelling was sensitive to only $f \sigma_8$, we would not expect redshift-evolution effects to be very relevant given that the change in $f \sigma_8$ within each redshift bin is expected to be very small, of order $\sim 0.01$, smaller than the observational uncertainties.

To test this line of argument explicitly, we performed a mock test for the $0.30 < z \leq 0.43$ sample using $\alpha = 0.73$. However, instead of constructing the mock galaxy population from the $z =0.4$ snapshot of the UNIT simulations, we derive it from the $z = 0.25$ snapshot. This simulates a mismatch between the redshift of the data $z = 0.25$ and the redshift assumed for the modelling, $z = 0.39$. We infer $f \sigma_8 = 0.464 \pm 0.021$ which is lower by $0.004$ than the mock constraint based on the comparable mock coming from the $z = 0.4$ snapshot. This shift to slightly lower $f \sigma_8$ is expected given that $f \sigma_8$ reduces by $0.007$ going from $z = 0.4$ to $z = 0.25$ for the cosmology of the UNIT simulation. We conclude that redshift evolution effects are likely only relevant to the extent that $f \sigma_8$ changes within a redshift bin. Since those are negligible compared to our observational uncertainties, redshift evolution inside each redshift bin likely has an insignificant effect on our results.

\subsection{Effects beyond SHAM mocks}

The test performed here are a first, highly non-trivial check that unbiased cosmological constraints can be obtained from the all-scale analysis of RSD data and goes beyond the results presented in \cite{Zhai2019_ApJ_874_95} and \cite{Lange2019_MNRAS_490_1870}. However, our analysis does not conclusively prove that unbiased cosmological constraints can always be achieved. Ultimately, further tests should be performed on a larger and more diverse sample of mock catalogues, ideally extracted from hydrodynamical simulations. In the following, we describe several effects that should be tested more thoroughly in the future.

\subsubsection{Baryonic feedback}

Since our modelling is based on dark matter-only simulations, we do not directly model the impact of baryonic feedback processes. For example, baryonic feedback is known to affect the matter distribution on small scales, particularly within a dark matter halo \citep{Jing2006_ApJ_640_119, Rudd2008_ApJ_672_19, vanDaalen2011_MNRAS_415_3649, Chisari2018_MNRAS_480_3962, VillaescusaNavarro2020_arXiv_2010_0619}. Since we marginalise over central and satellite phase-space parameters, some of the impact of baryonic feedback might be modelled through this approach.

The impact of baryons on the redshift-space clustering was discussed in \cite{Hellwing2016_MNRAS_461_11}. Particularly, the authors show that in the EAGLE hydrodynamical simulation, the impact of baryons on the dark matter redshift-space monopole and quadrupole power spectra can be as large as $\sim 4\%$ up to $k = 20 \, \hmpc$. However, the effect is primarily due to the suppression of the \textit{Fingers of God} effect, something we implicitly model via the parameter $\alpha_s$. At the same time, \cite{Hellwing2016_MNRAS_461_11} show that the impact of baryons on the peculiar velocities of dark matter haloes hosting massive galaxies, $M_\star > 3.5 \times 10^{10} \, M_\odot$ is much weaker. Particularly, for matched haloes in the dark matter-only and the hydrodynamical run, the velocity offset is consistent with zero and the $1 \sigma$ scatter between the peculiar velocities is $\sim 7 \mathrm{km} \, \mathrm{s}^{-1}$.

We conclude that baryons are unlikely to impact our conclusions in this work at a significant level. However, the back-reaction of baryons on the dark matter density and velocity field varies widely between different implementations of baryonic feedback \citep{Springel2018_MNRAS_475_676}. A detailed study of baryonic feedback on cosmological constraints from small-scale redshift space distortions is beyond the scope of this paper but warranted for future studies that use even higher precision measurements.

\subsubsection{Galaxy--halo connection}

Our modelling is based on a complex galaxy--halo model that is tested on non-trivial mock catalogues. Nonetheless, it is, in principle, possible that our parametrised galaxy--halo model is not flexible enough to yield unbiased cosmology constraints from more complex galaxy models or real data.

For example, both our HOD model and our SHAM mock implement galaxy assembly bias through a correlation of galaxy number with $V_{\rm max}$. However, \cite{Hadzhiyska2020_MNRAS_493_5506} and \cite{Xu2021_MNRAS_tmp_129} have recently shown in theoretical models of galaxy formation that $V_{\rm max}$ is insufficient to capture the full galaxy assembly bias effect. Similarly, \cite{Yuan2020_arXiv_2010_4182} argue that assembly bias with respect to both concentration and large-scale over-density is needed to get a good fit to the redshift-space clustering of BOSS CMASS galaxies. However, even if overdensity-based assembly bias was needed to model the data, it is not clear that this biases $f \sigma_8$ constraints significantly since neglecting $V_{\rm max}$-based assembly bias does not significantly alter the $f \sigma_8$ posterior constraints \citep{Lange2019_MNRAS_490_1870}. Additionally, as described in section \ref{sec:results}, we obtain a very good fit to the data without over-density-based assembly bias, contrary to \cite{Yuan2020_arXiv_2010_4182}.

\section{Application to observations}
\label{sec:results}

\begin{figure*}
    \centering
    \subfloat{\includegraphics{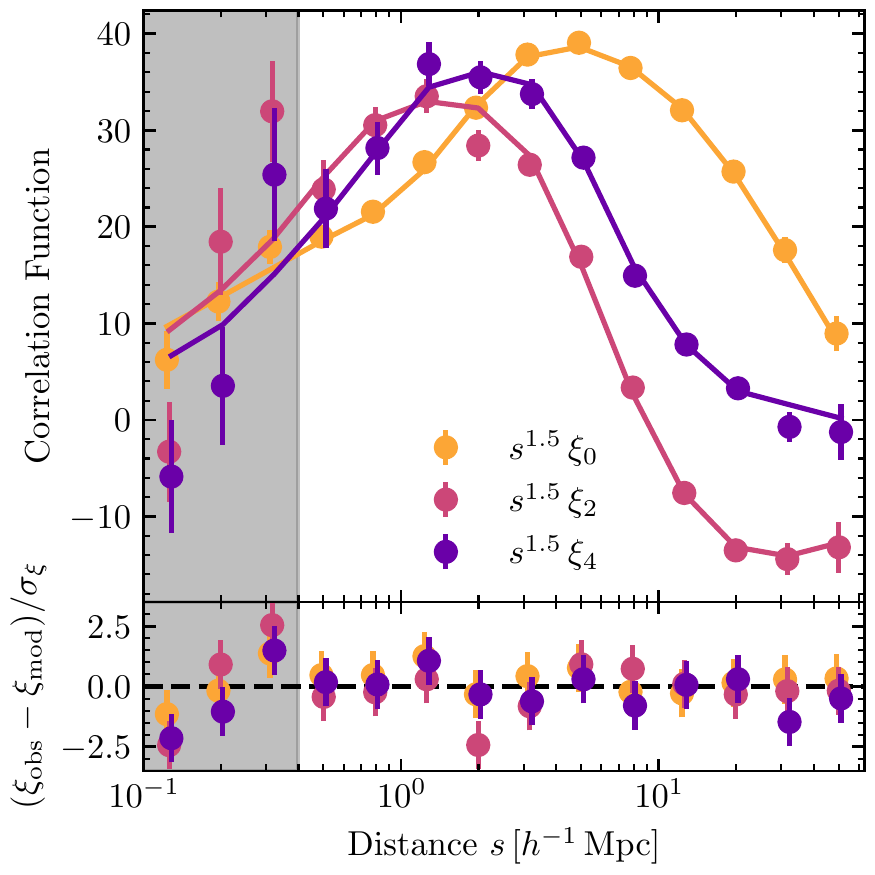}}
    \subfloat{\includegraphics{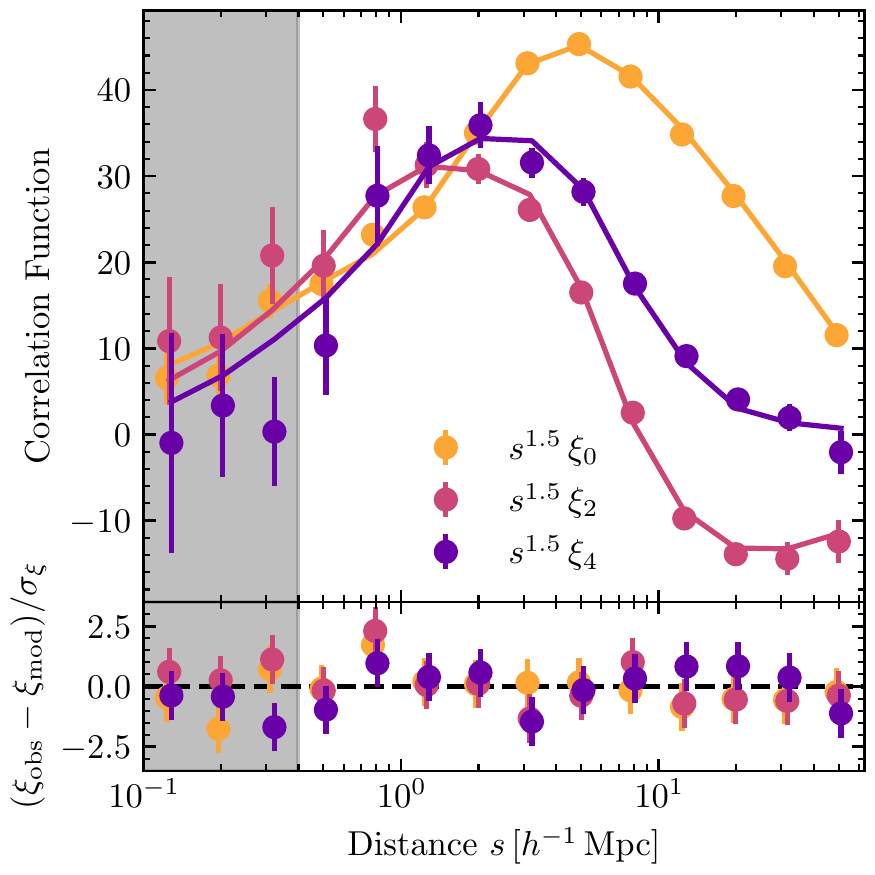}}
    \caption{Measurements of the monopole (blue), quadrupole (purple) and hexadecapole (orange) moments of the redshift-space correlation function. The left-hand side shows the measurements for the $z = 0.25$ sample and the right-hand side the observations for the $z = 0.40$ sample. All quantities are expressed in units of $\hmpc$. We also show as solid lines the best-fitting HOD model from the Aemulus simulation suite. The lower panels show the difference between the observations and the best-fit models in units of the observational uncertainty. Scales below $s = 400 \, \hkpc$ are excluded from the fit.}
    \label{fig:measurements}
\end{figure*}

After performing successful tests on mock catalogues, we now model the actual observations. In Figure~\ref{fig:measurements}, we show the measurements of the multipole moments of the redshift-space correlation functions in the two redshift bins. We also show as solid lines the respective best-fit models.

\subsection{Goodness-of-fit}

Using scales from $400 \, \hkpc$ to $63 \, \hmpc$, we find $\chi^2 = 20.9$ and $21.5$ for the $z = 0.25$ and $z = 0.40$ samples, respectively. The number of measurements points is $34$. Additionally, we have $11$ galaxy--halo connection parameters and $7$ cosmological parameters. However, the number of degrees of freedom of the model is likely smaller than $11 + 7$. Most importantly, we argued in section~\ref{sec:mock_tests}, that there is effectively only one cosmological parameter determining the fit such that the effective number of degree of freedom of the model is $\lesssim 12$. Hence, the number of degree of freedom of the fit is $\gtrsim 22$, in which case the $\chi^2$ values indicate a good fit of the model to the data.

\subsection{Constraints on \texorpdfstring{$f \sigma_8$}{f sigma8}}

We calculate the cosmological evidence with respect to the two observations for all $40$ simulations of the Aemulus simulation suite. The results are listed in appendix \ref{section:simulation_statistics}, thereby allowing the reader to derive cosmological constraints for different models for the evidence $\mathcal{Z} (\mathbf{D} | \mathcal{C})$.

\begin{figure*}
    \centering
    \subfloat{\includegraphics{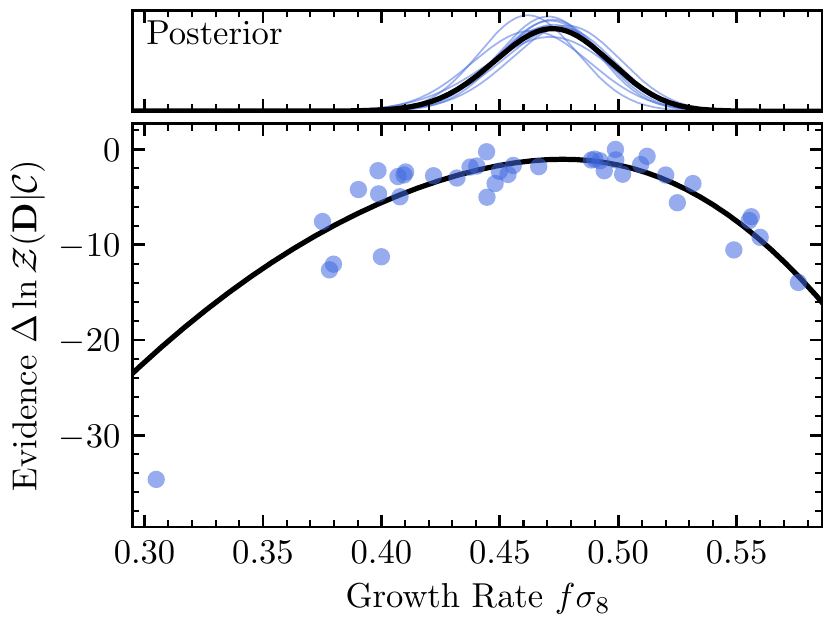}}
    \subfloat{\includegraphics{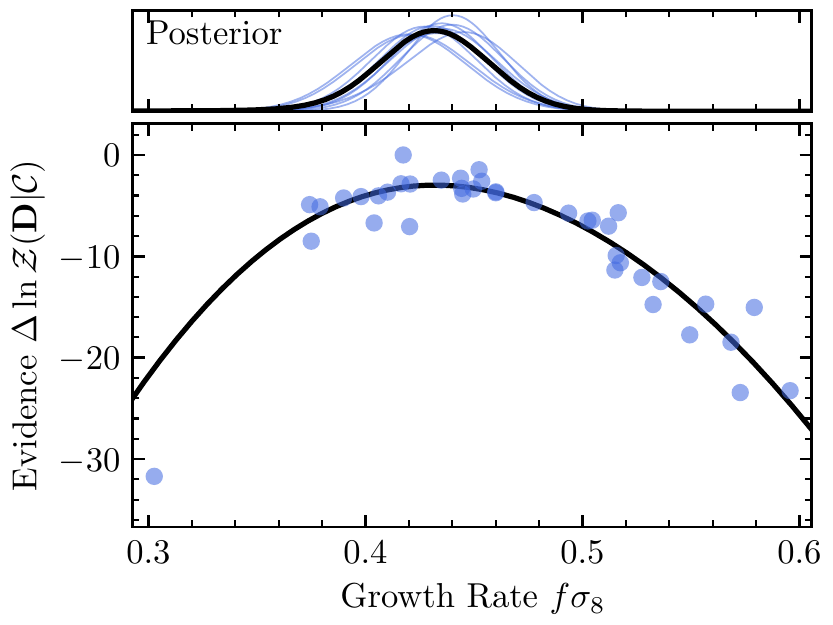}}
    \caption{Cosmological evidence as a function of $f \sigma_8$ for the BOSS LOWZ data. The left-hand panel shows the results for the $z = 0.25$ sample and the right-hand side the findings for the $z = 0.40$ sample. Blue points in the lower panels correspond to each of the $40$ regular Aemulus simulations. Similarly, the black line is the best-fit model to the evidence. The upper panels show draws from the posterior of the evidence model as thin blue lines and the total effective $f \sigma_8$ posteriors as the black lines.}
    \label{fig:log_ev_fs8}
\end{figure*}

In Figure~\ref{fig:log_ev_fs8}, we show the distribution of $\ln \mathcal{Z}$ as a function of $f \sigma_8$ for both observations and all simulations. In the same figure, we also show the derived constraints on $f \sigma_8$ if the evidence is assumed to only depend on $f \sigma_8$. We find $f \sigma_8 = 0.471 \pm 0.024$ and $0.431 \pm 0.025$ for redshifts $0.25$ and $0.40$, respectively. Similar to section~\ref{sec:mock_tests}, we find that the $f \sigma_8$ constraints are virtually unchanged if the evidence is assumed to depend on additional cosmological parameters like $w_0$.

\subsection{Robustness to analysis choices}
\label{subsec:robustness}

\begin{table*}
    \centering
    \begin{tabular}{c|c|ccc|ccc}
    Method & $N_{\rm data}$ & $f \sigma_8$ & $\chi^2$ & $\chi^2 / {\rm dof}$ & $f \sigma_8$ & $\chi^2$ & $\chi^2 / {\rm dof}$\\
    && \multicolumn{3}{c|}{$0.18 < z \leq 0.3$}& \multicolumn{3}{c}{$0.3 < z \leq 0.43$}\\
    \hline\hline
    Default & 34 & $0.471 \pm 0.024$ & $20.9$ & $0.95$ & $0.431 \pm 0.026$ & $21.5$ & $0.98$ \\
    no $\xi_4$ & 23 & $0.469 \pm 0.027$ & $11.5$ & $1.04$ & $0.436 \pm 0.023$ & $9.4$ & $0.86$ \\
    Hartlap & 34 & $0.464 \pm 0.022$ & $18.2$ & $0.83$ & $0.427 \pm 0.026$ & $16.9$ & $0.77$ \\
    with $w_{\rm p}$ & 45 & $0.450 \pm 0.021$ & $44.3$ & $1.34$ & $0.428 \pm 0.025$ & $35.7$ & $1.08$ \\
    $s > 1.0 \, h^{-1} \, \mathrm{Mpc}$ & 28 & $0.473 \pm 0.023$ & $16.6$ & $1.04$ & $0.438 \pm 0.024$ & $11.2$ & $0.70$ \\
    $s > 2.5 \, h^{-1} \, \mathrm{Mpc}$ & 22 & $0.495 \pm 0.030$ & $6.1$ & $0.61$ & $0.445 \pm 0.024$ & $5.7$ & $0.57$ \\
    $s > 6.3 \, h^{-1} \, \mathrm{Mpc}$ & 16 & $0.542 \pm 0.042$ & $3.3$ & $0.81$ & $0.469 \pm 0.039$ & $2.4$ & $0.61$ \\
    $s < 25 \, h^{-1} \, \mathrm{Mpc}$ & 28 & $0.462 \pm 0.026$ & $17.8$ & $1.11$ & $0.446 \pm 0.028$ & $17.9$ & $1.12$ \\
    $s < 10 \, h^{-1} \, \mathrm{Mpc}$ & 22 & $0.415 \pm 0.028$ & $12.1$ & $1.21$ & $0.406 \pm 0.034$ & $10.7$ & $1.07$ \\
    \end{tabular}
    \caption{Dependence of the cosmological constraints to variations in the analysis choices. In order, we show the results for the default choices, when excluding the hexadecapole, when using the \protect\cite{Hartlap2007_AA_464_399} correction to the covariance matrix, when including the projected correlation function and when applying a variety of different scale cuts. For the degrees of freedom, $\rm dof$, we set ${\rm dof} = N_{\rm data} - 12$, where $N_{\rm data}$ is the number of data points and $12$ is the number of free parameters in the model, i.e. $11$ galaxy--halo connection parameters and one effective cosmological parameter, $f \sigma_8$.}
    \label{tab:fs8_modes}
\end{table*}

In the following, we will analyse how our constraints on $f \sigma_8$ change if we apply reasonable variations to the default analysis. All results are tabulated in Table~\ref{tab:fs8_modes}.

\subsubsection{Scale dependence}

\begin{figure*}
    \centering
    \subfloat{\includegraphics{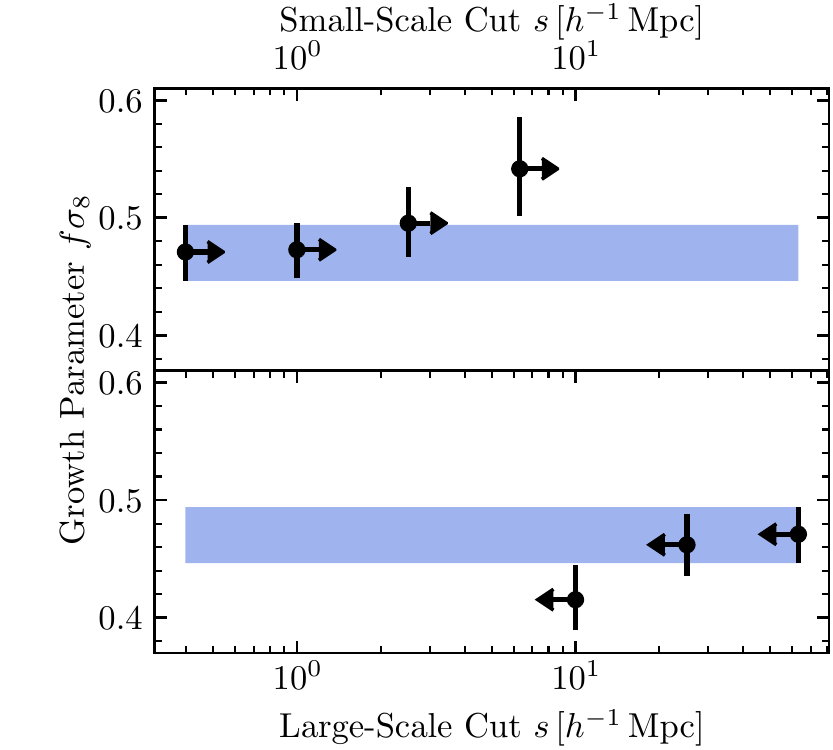}}
    \subfloat{\includegraphics{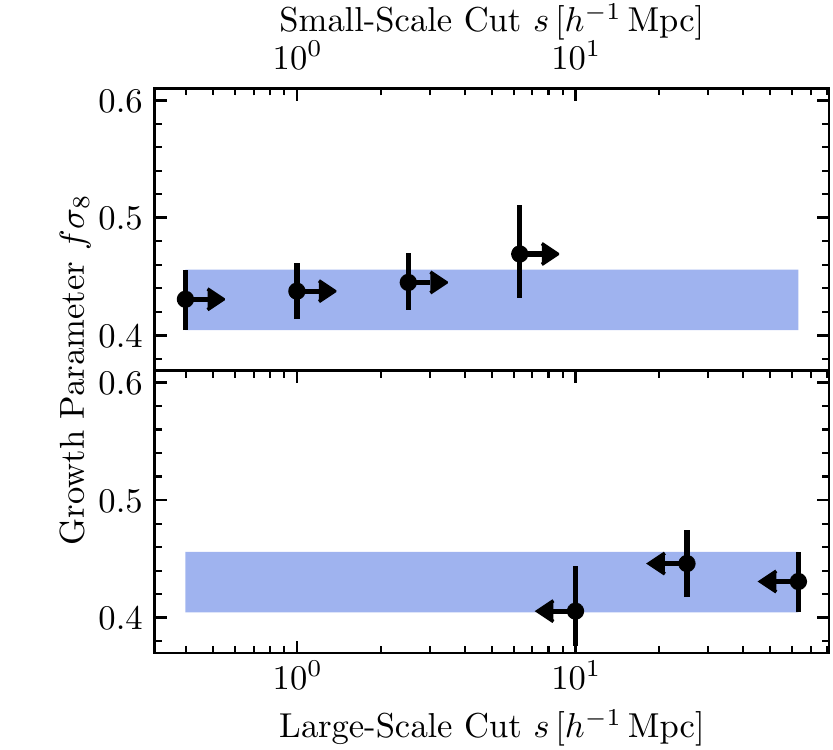}}
    \caption{The dependence of the cosmological constraints on the scales considered in the analysis. By default, we utilize scales within $400 \, \hkpc < s < 63 \, \hmpc$. We show the results for the $0.18 < z \leq 0.3$ (left) and the $0.3 < z \leq 0.43$ (right) samples. Upper panels indicate how constraints on $f \sigma_8$ change if the minimum scale considered is raised from its original $s_{\rm min} = 400 \, \hkpc$ value. Similarly, lower panels show results for reducing the maximum scale from $63 \, \hmpc$ to a lower value. The blue rectangle in each panel indicates the default scale cuts and constraints.}
    \label{fig:scale_cuts_consistency}
\end{figure*}

By default, we use scales within $400 \, \hkpc < s < 63 \, \hmpc$ in our analysis. Here, we study how different scale cuts, both on the lower and the upper limit in $s$, affect our cosmological constraints. In Figure~\ref{fig:scale_cuts_consistency}, we show the constraints on $f \sigma_8$ for the two galaxy samples and different scales cuts.

For the low-redshift sample, excluding smaller scales shifts the posterior constraint on $f \sigma_8$ upward, reaching $0.541 \pm 0.043$ when excluding scales below $6.3 \hmpc$. Similarly, excluding large scales drives $f \sigma_8$ down, reaching $0.416 \pm 0.027$ when limiting the analysis to scales below $10 \hmpc$. That small and large-scale cuts drive the constraints in opposite direction seems to be a consequence of small scales ($< 10 \hmpc$) preferring lower values for $f \sigma_8$ than larger scales. The $>6.3 \hmpc$ and the $<10.0 \hmpc$ samples, with the exception of $6.3 \hmpc < s < 10 \hmpc$, use completely different scales. Here, we find a $\sim 2.5 \sigma$ discrepancy between the two scale cuts for the $0.18 < z \leq 0.3$ galaxy sample. We also observe a similar albeit weaker behaviour for the $0.3 < z \leq 0.43$ sample. While the smaller scales also prefer lower values for $f \sigma_8$, the discrepancy is only of the order of $\sim 1 \sigma$. Additionally, we do not observe comparable trends in the mock tests in section \ref{sec:mock_tests} for any of the five mock samples analysed. These trends are discussed further in the discussion section.

\subsubsection{Impact of hexadecapole}

By default, we use the monopole $\xi_0$, quadrupole $\xi_2$ and hexadecapole $\xi_4$ of the redshift-space correlation function as observables. We repeat the analysis here without the hexadecapole $\xi_4$ to see how sensitive our analysis is with respect to this observable. We now find $f \sigma_8 = 0.469 \pm 0.027$ and $0.436 \pm 0.023$ for the two samples, indicating only a negligible shift and little impact on the accuracy of the constraints. Generally, only fitting the monopole and quadrupole moments reliably predicts the hexadecapole. Thus, including $\xi_4$ has little impact on the cosmological constraints.

\subsubsection{Dependence on covariance matrix}

Our default analysis is performed using the covariance matrix described in section~\ref{subsec:covariance}. This covariance matrix uses a physically motivated way to correct for biases in the covariance matrix estimate due to a finite number of jackknife samples. A commonly used alternative is the \cite{Hartlap2007_AA_464_399} correction,
\begin{equation}
    \hat{\Phi} = \frac{N_{\rm jackknife} - N_{\rm data} - 2}{n_{\rm jackknife} - 1} \Sigma^{-1},
\end{equation}
where $\hat{\Phi}$ is used as the precision matrix instead of $\Sigma^{-1}$ in equation~(\ref{eq:likelihood}). In the above equation $n_{\rm jackknife} = 74$ is the number of jackknife samples and $N_{\rm data} = 34$ the number of measurements.

We have repeated the entire analysis using the \cite{Hartlap2007_AA_464_399} correction but find very similar results. For the $z = 0.25$ and $z = 0.40$ samples we now find $f \sigma_8 = 0.464 \pm 0.022$ and $0.427 \pm 0.026$, respectively. Thus, our results are not significantly affected by the choice of covariance matrix. Finally, the best-fit $\chi^2$ values improve from $20.9$ and $21.5$ to $18.2$ and $16.9$ for the two samples, respectively.

\subsubsection{Projected correlation function}

Previous studies \citep[e.g.][]{Reid2014_MNRAS_444_476, Guo2015_MNRAS_446_578, Zhai2019_ApJ_874_95} of RSDs have used the so-called projected correlation function as an observable. The projected correlation function $w_{\rm p}$ is defined via
\begin{equation}
    w_{\rm p} (r_{\rm p}) = \int\limits_{-\pi_{\rm max}}^{+\pi_{\rm max}} \xi(r_{\rm p}, \pi) d \pi,
\end{equation}
Our rationale for not using this observables is that it mostly encodes data that is already contained within the multipoles. Additionally, building a covariance matrix becomes more difficult with increasing number of data points. Nonetheless, we test here how the inclusion of $w_{\rm p}$ affects our cosmological constraints. We use the same scales for $r_{\rm p}$ as we used for $s$ thus far. \footnote{In the following, we use the same smoothing technique described in section~\ref{subsec:covariance} to estimate the covariance matrix. Note that for the covariance between $w_{\rm p}$ and any multipole moment, we do not exclude elements where $r_{\rm p} = s$ from the smoothing. The reason is that $w_{\rm p} (r_{\rm p})$ depends on the correlation function at all $r_{\rm p} \leq s \leq \left( r_{\rm p}^2 + \pi_{\rm max}^2 \right)^{1/2}$, not just $s = r_{\rm p}$.}

We find that for the $0.3 < z \leq 0.43$ sample, the inclusion of the projected correlation function only shifts the $f \sigma_8$ constraint by $0.004$ and does not significantly affect the uncertainty. However, for the $0.18 < z \leq 0.3$ galaxy sample, the impact is stronger as the constraint shifts from $0.471 \pm 0.024$ to $0.450 \pm 0.021$. Finally, we note that the inclusion of $11$ data points of the projected correlation function increases the best-fit $\chi^2$ by $23.4$ and $14.2$ for the low and the high-redshift sample, respectively. For the low-redshift sample, this might indicate a slight tension of $w_{\rm p}$ and the multipoles under our theoretical models, though we defer a detailed investigation of this and potentially related issues to a future analysis based on DESI data (see \S\ref{subsec:consistency_scales} for further discussion).

\subsection{Galaxy--halo connection}

In the following, we will present constraints on galaxy--halo connection parameters $\mathcal{G}$ obtained after marginalising over cosmology. Formally, the posterior constraint on $\mathcal{G}$ can be computed via
\begin{equation}
    P (\mathcal{G} | \mathbf{D}) = \int P(\mathcal{G}, \mathcal{C} | \mathbf{D}) {\rm d} \mathcal{C} = \int P(\mathcal{G} | \mathcal{C}, \mathbf{D}) P(\mathcal{C} | \mathbf{D}) {\rm d} \mathcal{C}.
\end{equation}
Since the simulations sample the prior space in $\mathcal{C}$ and the posterior probability of $\mathcal{C}$ is proportional to the cosmological evidence, an estimate for the posterior probability of $\mathcal{G}$ marginalised over $\mathcal{C}$ is
\begin{equation}
    \widehat{P (\mathcal{G} | \mathbf{D})} = \sum\limits_i \hat{\mathcal{Z}} (\mathcal{C}_i | \mathbf{D}) P(\mathcal{G} | \mathcal{C}_i, \mathbf{D})\,,
    \label{eq:posterior}
\end{equation}
where the sum goes over all of the Aemulus simulations. The posterior probability of galaxy--halo connection parameters at a given cosmology, $P(\mathcal{G} | \mathcal{C}_i, \mathbf{D})$, is a natural by-product of the {\sc MultiNest} analysis. Note that we use the model evidence $\hat{\mathcal{Z}}$ instead of the directly measured evidence $\mathcal{Z}$ because the latter is more noisy, as can be seen in Figure~\ref{fig:log_ev_fs8}.

\begin{table}
    \centering
    \begin{tabular}{c|cc}
        \multirow{2}{*}{Parameter} & \multicolumn{2}{c}{Posterior} \\
        & $0.18 < z \leq 0.30$ & $0.30 < z \leq 0.43$\\
        \hline
        $\log M_{\rm min}$ & $13.161_{-0.054}^{+0.116}$ & $13.45_{-0.17}^{+0.17}$ \\
        $\log M_0$ & $12.41_{-0.29}^{+0.37}$ & $12.41_{-0.29}^{+0.39}$ \\
        $\log M_1$ & $14.339_{-0.046}^{+0.057}$ & $14.547_{-0.095}^{+0.181}$ \\
        $\sigma_{\log M}$ & $0.25_{-0.11}^{+0.20}$ & $0.47_{-0.21}^{+0.15}$ \\
        $\alpha$ & $1.32_{-0.13}^{+0.11}$ & $1.35_{-0.28}^{+0.34}$ \\
        $f_\Gamma$ & $0.756_{-0.069}^{+0.115}$ & $0.78_{-0.18}^{+0.16}$ \\
        $A_{\rm cen}$ & $0.03_{-0.39}^{+0.45}$ & $0.38_{-0.34}^{+0.33}$ \\
        $A_{\rm sat}$ & $-0.43_{-0.35}^{+0.42}$ & $-0.50_{-0.38}^{+0.96}$ \\
        $\log \eta$ & $0.01_{-0.12}^{+0.12}$ & $-0.21_{-0.17}^{+0.18}$ \\
        $\alpha_{\rm c}$ & $0.145_{-0.090}^{+0.071}$ & $0.238_{-0.081}^{+0.082}$ \\
        $\alpha_{\rm s}$ & $0.927_{-0.079}^{+0.112}$ & $0.98_{-0.12}^{+0.14}$ \\
    \end{tabular}
    \caption{Posterior constraints on galaxy--halo connection parameters after marginalising over cosmology.}
    \label{tab:posterior_galaxy_halo_connection}
\end{table}

\begin{figure*}
    \centering
    \includegraphics{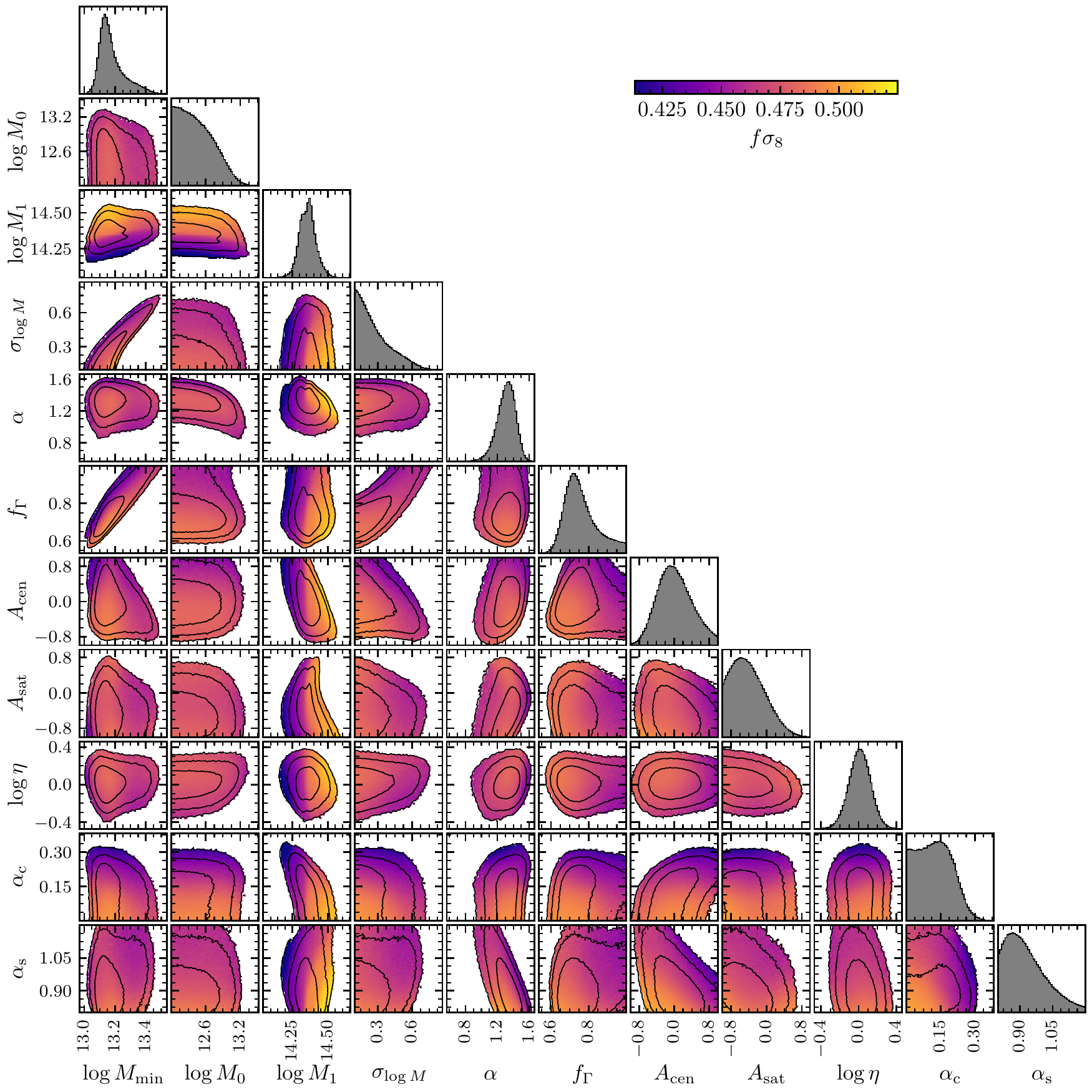}
    \caption{Posterior constraints on galaxy--halo connection parameters for the $0.18 < z \leq 0.30$ sample after marginalisation over cosmology. Panels along the diagonal show marginalised one-dimensional posteriors. The off-diagonal panels show as contours the $68\%$, $95\%$ and $99\%$ posterior mass. Additionally, background colours indicate the marginalised mean of the $f \sigma_8$ posterior as a function of galaxy--halo connection parameter.}
    \label{fig:corner_1}
\end{figure*}

\begin{figure*}
    \centering
    \includegraphics{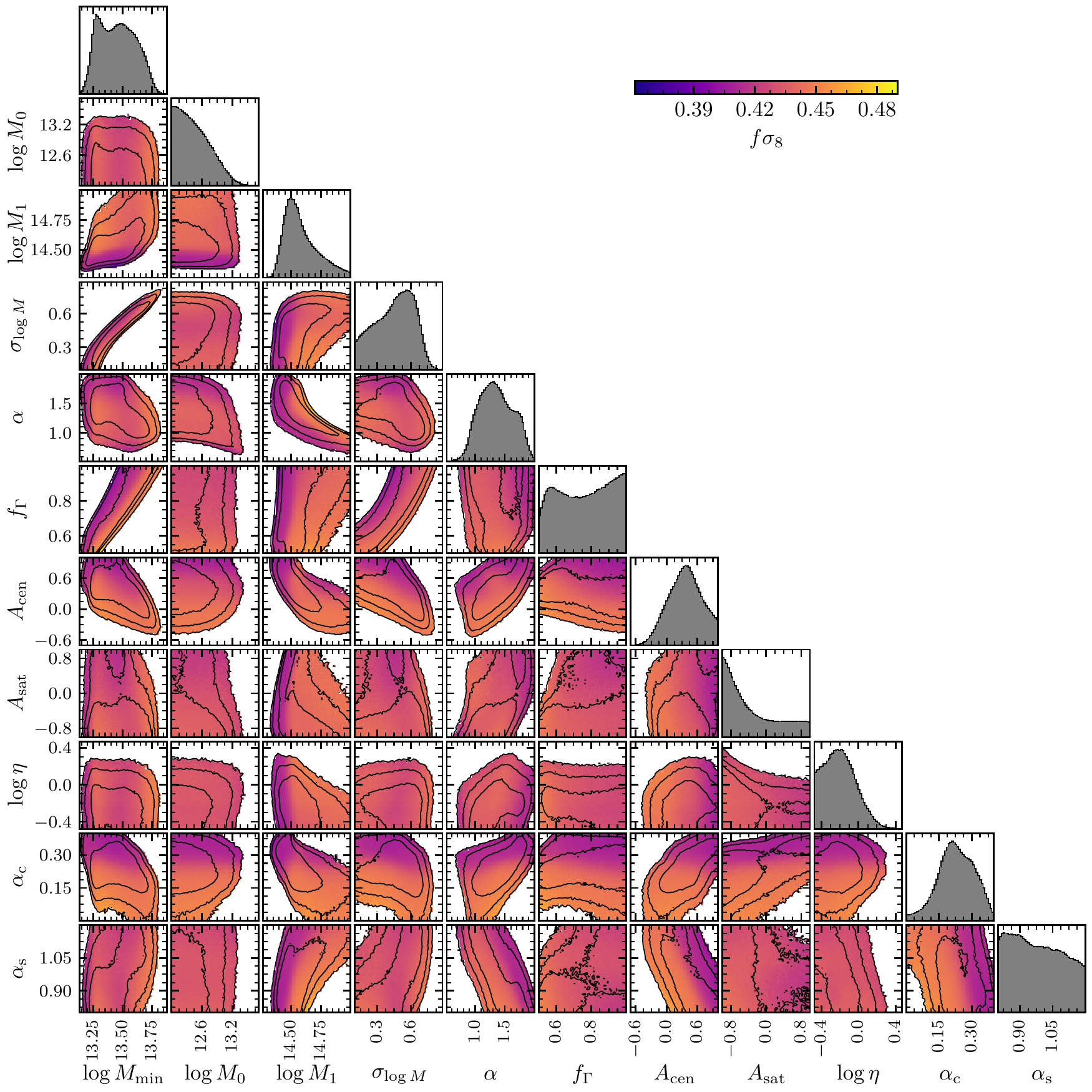}
    \caption{Similar to Figure~\ref{fig:corner_1} but for the $0.30 < z \leq 0.43$ sample.}
    \label{fig:corner_2}
\end{figure*}

In Table~\ref{tab:posterior_galaxy_halo_connection} and Figures \ref{fig:corner_1} and \ref{fig:corner_2}, we show our derived posterior constraints on the galaxy--halo connection parameters. In the following, we discuss some noteworthy findings. First, we do not find strong evidence for assembly bias, i.e. $A_{\rm cen} \neq 0$ or $A_{\rm sat} \neq 0$. However, we only place very weak constraints, in part due to the degeneracy between assembly bias and $f \sigma_8$ \citep{Lange2019_MNRAS_490_1870}. Furthermore, our findings are consistent with satellites following an NFW profile with the same concentration as dark matter, i.e. $\eta = 1$. Additionally, we find some evidence of central velocity bias $\alpha_c > 0$ for the $0.30 < z \leq 0.43$ sample but no evidence for central velocity bias for the $0.18 < z \leq 0.30$ sample. Finally, our analysis is also consistent with no satellite velocity bias, i.e. $\alpha_s = 1$, for both samples.

Figures \ref{fig:corner_1} and \ref{fig:corner_2} also show the degeneracy between galaxy--halo connection parameters and $f \sigma_8$. At each point in the galaxy--halo connection parameter space, we calculate the average $f \sigma_8$ of the simulation boxes contributing to the posterior mass. We find that $\log M_1$, $A_{\rm cen}$ and $\alpha_c$ are most strongly degenerate with the $f \sigma_8$ posterior. Finally, we note that some of our galaxy--halo connection parameters are limited by the prior. In many cases, the limiting priors are determined by the ranges in which the parameters can be defined, e.g. $f_\Gamma \leq +1$, $-1 \leq | A_{\rm cen} | \leq +1$, $-1 \leq | A_{\rm sat} | \leq +1$ and $\alpha_{\rm c} > 0$. Additionally, $\log M_0$ has little impact on the HOD once $\log M_0 < 12$ and $\sigma_{\log M}$ describing the rate of transition from $N_{\rm cen} = 0$ to $N_{\rm cen} = 1$ is not expected to be very narrow, i.e. $0.0 \leq \sigma_{\log M} < 0.1$. Given that $M_0$ and $\sigma_{\log M}$ do not strongly correlate with $f \sigma_8$, it is unlikely that our prior choice impacts the $f \sigma_8$ posterior.

\subsection{Comparison with galaxy-galaxy lensing}

Under cosmological $\Lambda$CDM parameters compatible with the \cite{PlanckCollaboration2020_AA_641_6} CMB analysis, \cite{Leauthaud2017_MNRAS_467_3024} have shown that several galaxy--halo connection models fitted to the clustering properties of galaxies in BOSS do not correctly predict their galaxy-galaxy lensing amplitude. The lensing amplitude, the so-called excess surface density $\Delta\Sigma$, is defined as
\begin{equation}
    \Delta\Sigma(r_{\rm p}) = \langle \Sigma(<r_{\rm p}) \rangle - \Sigma(r_{\rm p}) \, .
\end{equation}
In the above equation $r_{\rm p}$ is the projected comoving separation from the galaxy, $\langle \Sigma(<r) \rangle$ the mean surface mass density (in comoving units) for separations less than $r_{\rm p}$ and $\Sigma(r_{\rm p})$ the surface mass density at $r_{\rm p}$. The combination of galaxy clustering and galaxy-galaxy lensing is particularly sensitive to the cosmological parameters $\Omega_{\rm m, 0}$ and $\sigma_8 (z)$ \citep[see e.g.][]{More2013_MNRAS_430_747, Wibking2019_MNRAS_484_989}. Thus, a mismatch in the lensing prediction after fitting the clustering and assuming \cite{PlanckCollaboration2020_AA_641_6} parameters could indicate a tension in these two parameters with respect to the CMB under $\Lambda$CDM. At the same time, especially on small scales, care must be taken because other effects like baryonic feedback \citep{Leauthaud2017_MNRAS_467_3024, Lange2019_MNRAS_488_5771} or galaxy assembly bias \citep{Lange2019_MNRAS_488_5771, Yuan2019_MNRAS_486_708, Yuan2020_MNRAS_493_5551} can affect the lensing prediction. Similarly, modifications to general relativity could also modify the expected lensing amplitude \citep{Leauthaud2017_MNRAS_467_3024}.

\begin{figure*}
    \centering
    \subfloat{\includegraphics{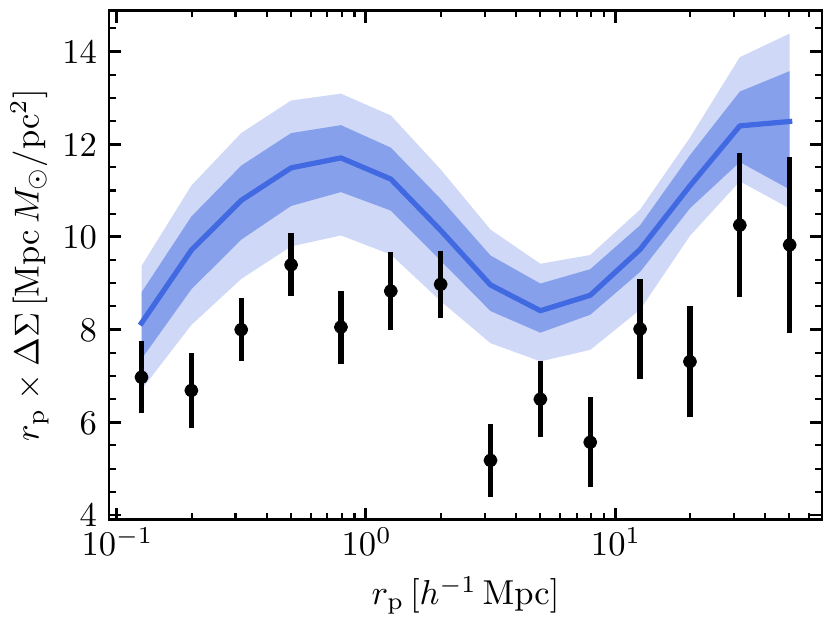}}
    \subfloat{\includegraphics{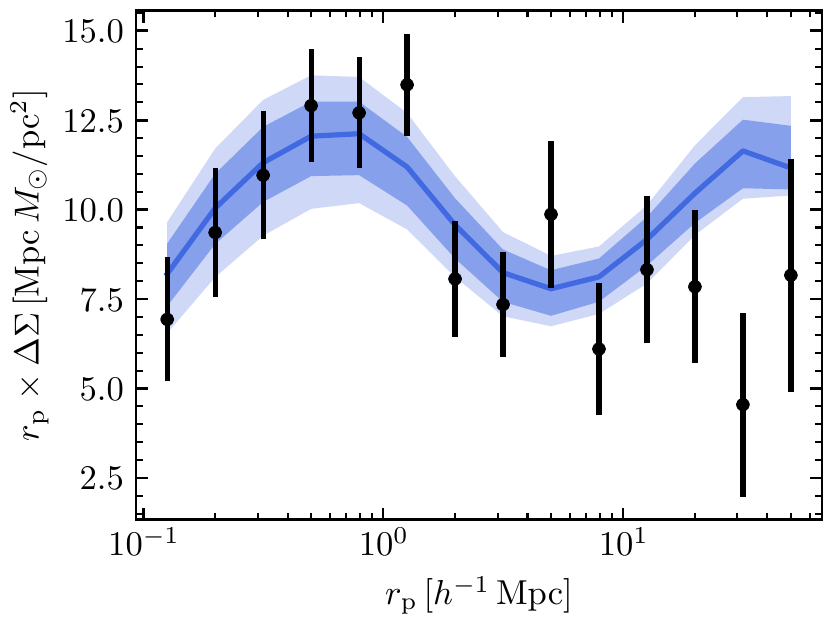}}
    \caption{Posterior predictions for the galaxy-galaxy lensing amplitude of galaxies in the $0.18 < z \leq 0.30$ (left) and the $0.30 < z \leq 0.43$ sample (right). Bands denote the $68\%$ and $95\%$ posterior mass. Measurements from cross-correlating BOSS LOWZ targets with shape catalogues from SDSS are shown by error bars. The lensing amplitude is over-predicted for the low-redshift sample. Our prediction agrees much better with the measurements for the high-redshift sample but in this redshift range the inferred $f \sigma_8$ from RSDs is $10\%$ lower than the \protect\cite{PlanckCollaboration2020_AA_641_6} $\Lambda$CDM prediction at $2 \sigma$ significance.}
    \label{fig:lensing_predictions}
\end{figure*}

We compute $\Delta\Sigma$ for a given mock catalogue using a random $0.5\%$ subset of all simulation particles and the \texttt{mean\_delta\_sigma} routine in {\sc halotools}. In Figure~\ref{fig:lensing_predictions} we show the marginalised posterior predictions for the galaxy-galaxy lensing amplitudes of the two samples. The marginalisation over cosmology is performed analogously to equation (\ref{eq:posterior}). We also show measurements of the galaxy-galaxy lensing amplitude from cross-correlating BOSS LOWZ targets with sources from SDSS. We refer the reader to \cite{Singh2020_MNRAS_491_51} regarding details of the SDSS lensing measurements. Note that there is an overall $6\%$ uncertainty on the total normalization of the lensing amplitude, stemming from photometric redshift uncertainties in SDSS. 

In agreement with previous studies \citep{Leauthaud2017_MNRAS_467_3024, Lange2019_MNRAS_488_5771, Yuan2020_MNRAS_493_5551, Lange2020_arXiv_2011_2377}, we find that the lensing amplitude tends to be over-predicted. The over-prediction is significant for the low-redshift sample while our prediction agrees much better with the measurements for the high-redshift sample. However, for the higher redshift sample, the inferred $f \sigma_8$ is $10\%$ from RSDs is lower than the \cite{PlanckCollaboration2020_AA_641_6} $\Lambda$CDM prediction at $2 \sigma$ significance.

The novelty of the result here is that the lensing prediction is marginalised over cosmology whereas the aforementioned studies showed the mismatch under $\Lambda$CDM parameters of the \cite{PlanckCollaboration2020_AA_641_6} CMB analysis. We note, however, that our marginalisation over cosmology implicitly includes the cosmological priors of the Aemulus simulations \citep{DeRose2019_ApJ_875_69}. Despite this additional freedom in cosmological parameters, we still find that our model for the low-redshift sample fails to correctly predict the galaxy-galaxy lensing amplitude. Finally, it is worth remembering that our lensing predictions, which are based on dark matter-only simulations, do not account for the impact of baryons on the matter distribution \citep{Leauthaud2017_MNRAS_467_3024, Lange2019_MNRAS_488_5771, Amodeo2020_arXiv_2009_5558}. Furthermore, \cite{Yuan2020_arXiv_2010_4182} show that overdensity-based assembly bias can further reduce this lensing tension, even when modelled in addition to concentration or $V_{\rm max}$-based assembly bias. Both effects, baryonic feedback and more complex assembly bias models, could potentially reconcile or at least alleviate the tension on small scales, $r_{\rm p} \lesssim 3 \, \hmpc$, but are unlikely to impact differences on larger scales.

\section{Discussion}
\label{sec:discussion}

\subsection{Constraints on the growth rate}

\begin{figure*}
    \centering
    \includegraphics{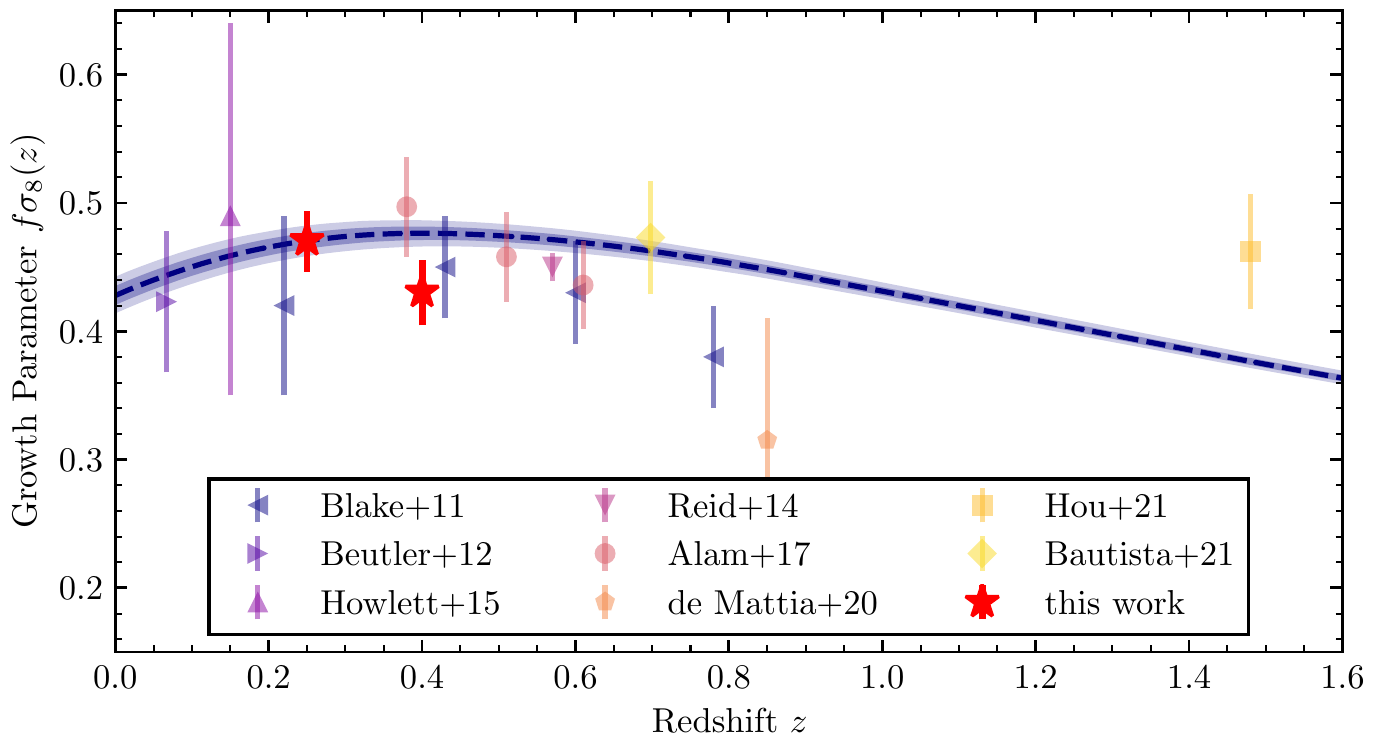}
    \caption{Comparison of different constraints on the growth rate of structure. We show results from the \protect\cite{PlanckCollaboration2020_AA_641_6} CMB analysis assuming a $\Lambda$CDM cosmology (blue). The bands denote the $1$ and $2\sigma$ ranges. Additionally, we show large-scale structure constraints from the WiggleZ survey \protect\citep{Blake2011_MNRAS_415_2876}, the 6dF Galaxy Survey \protect\citep{Beutler2012_MNRAS_423_3430} and the SDSS main galaxy sample \protect\citep{Howlett2015_MNRAS_449_848}, BOSS \protect\citep{Alam2017_MNRAS_470_2617} and eBOSS \protect\citep{deMattia2020_MNRAS_tmp_3648, Bautista2021_MNRAS_500_736, Hou2021_MNRAS_500_1201}. We also compare our results (red) against the small-scale BOSS CMASS RSD analysis of \protect\cite{Reid2014_MNRAS_444_476}. When necessary, some results are slightly shifted in the $x$-direction for clarity.}
    \label{fig:fs8}
\end{figure*}

In Figure~\ref{fig:fs8}, we compare our constraints on $f \sigma_8 (z)$ against other results from the literature and the $\Lambda$CDM \cite{PlanckCollaboration2020_AA_641_6} CMB prediction. The literature comparison is chosen to represent the leading results on large scales from the 6dF Galaxy Survey \citep{Beutler2012_MNRAS_423_3430}, the WiggleZ survey \citep{Blake2011_MNRAS_415_2876}, SDSS main galaxy sample \citep{Howlett2015_MNRAS_449_848}, BOSS \citep{Alam2017_MNRAS_470_2617} and eBOSS \citep{deMattia2020_MNRAS_tmp_3648, Bautista2021_MNRAS_500_736, Hou2021_MNRAS_500_1201}. Our results are consistent with other results in the literature as well as CMB $\Lambda$CDM predictions. At the same time, our results follow the trend of low-redshift measurements of $f \sigma_8$ falling slightly below the $\Lambda$CDM+CMB prediction. Future work is needed to verify whether this difference is statistically significant. We also see that our constraints on $f \sigma_8$ are roughly a factor of two more stringent than any other study on large scales, demonstrating the potentially large benefits of extending the analysis to the full range of scales accessible with observations.

In Figure~\ref{fig:fs8}, we also include the full-scale BOSS RSD analysis of \cite{Reid2014_MNRAS_444_476}. However, the analysis of \cite{Reid2014_MNRAS_444_476} is based on the assumption that changes in $f \sigma_8$ are completely degenerate with a simple linear scaling of the velocity field on all scales. \cite{Zhai2019_ApJ_874_95} argue that this approximation might lead to errors and cosmological parameter constraints that are artificially narrow. Our work is not based on this assumption and should therefore be more robust. We note, however, that the analysis of \cite{Reid2014_MNRAS_444_476} is based on $\sim 500,000$ galaxies in the BOSS CMASS sample. Given this roughly seven times larger galaxy sample, we estimate that if we applied our analysis framework to the same galaxy sample, we would obtain a roughly similar precision on $f \sigma_8$. Notwithstanding the validity of the scaling approximation in \cite{Reid2014_MNRAS_444_476}, it is encouraging to see that the non-linear analysis of \cite{Reid2014_MNRAS_444_476} shows results broadly comparable with ours: $f \sigma_8$ is compatible but slightly lower than the \cite{PlanckCollaboration2020_AA_641_6} $\Lambda$CDM prediction.

\subsection{Consistency between small and large scales}
\label{subsec:consistency_scales}

In Figure \ref{fig:scale_cuts_consistency} we showed how constraints on $f \sigma_8$ vary when excluding small or large scales $s$ from the analysis. Generally, we find consistent results between small and large scales, but there are also 1-2.5$\sigma$ hints of small scales ($< 10 \hmpc$) preferring lower values for $f \sigma_8$ than larger scales, especially for the low-redshift sample. When analysing mock catalogues, as described in section \ref{sec:mock_tests}, we did not find comparable trends of a strong scale dependence. There are a number of possibilities for the findings in the data. First, it could be a real trend in the data driven by some aspect of galaxy formation and/or baryonic physics. Future mock tests using more complex galaxy models than SHAM, ideally based on hydrodynamical simulations, are needed to investigate this possibility further. Second, it could be related to an effect beyond the traditional $\Lambda$CDM and $w$CDM models. For example, while modifications to general relativity like $f(R)$ gravity induce consistent changes to $f \sigma_8$ on large, linear scales, the impact has a strong scale dependence in the non-linear regime \citep{Fontanot2013_MNRAS_436_2672, He2018_NatAs_2_967, Alam2020_arXiv_2011_5771}. Thus, scale dependent $f \sigma_8$ constraints under $w$CDM might, in fact, be a sign of modifications to general relativity. Third, these trends could also be simply due to random statistical fluctuations. Upcoming surveys, especially DESI, will soon provide much larger samples of LRGs at similar redshifts. If the trends we see in the data are not due to random statistical flucations, higher signal to noise DESI data should easily confirm them with high significance.

\subsection{Constraining power as a function of scale}
\label{subsec:scale_dependent_constraints}

\begin{figure*}
    \centering
    \includegraphics{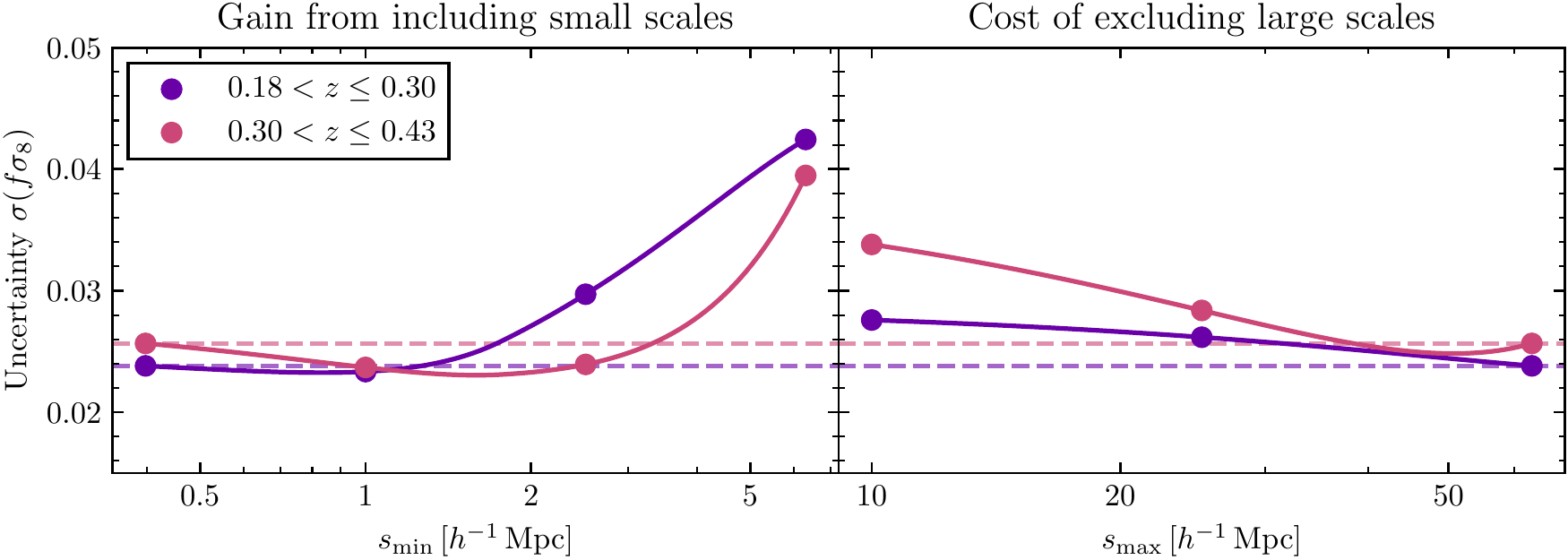}
    \caption{Similar to Figure~\ref{fig:scale_cuts_consistency} but showing how the constraints on $f \sigma_8$ depend on scale cuts. In the left panel, we show the dependence when the small-scale cut is increased compared to its default value of $s_{\rm min} = 400 \, \hkpc$. Similarly, the right panel shows the change in constraining power if the large-scale cut is reduced to values lower than $s_{\rm max} = 63 \, \hmpc$.}
    \label{fig:scale_cuts_precision}
\end{figure*}

The ability to harness the cosmological constraining power of the nonlinear regime is the driving motivation behind our development of the Cosmological Evidence Modelling technique, as well as our simulation-based forward modelling methodology. With Figure~\ref{fig:scale_cuts_precision} we highlight how our analysis reaps the benefits of this simulation and modelling effort, by showing how the precision of our cosmological constraints varies as a function of scales analysed. In both panels, our fiducial analysis is conducted over scales $s$ in the range $400 \, \hkpc < s < 63 \, \hmpc$. In the left-hand panel of Figure~\ref{fig:scale_cuts_precision}, the horizontal axis shows the minimum scale used in the analysis, $s_{\rm min};$ each point in the figure shows the results of an RSD analysis that includes information restricted to $s_{\rm min} < s < 63 \, \hmpc,$ with the y-axis showing the strength of the constraints on $f\sigma_8$ derived in each case. Thus larger values of $s_{\rm min}$ correspond to analyses which exclude more information from small scales, and so both curves curves in the left-hand panel naturally increase with $s_{\rm min.}$ 

There are two striking features about the left-panel of Figure~\ref{fig:scale_cuts_precision}. First, the constraining power improves dramatically as deeply nonlinear scales $s<10 \, \hmpc$ are included in the analysis. For example, an analysis with $s_{\rm min}=2.5 \, \hmpc$ has a $50\%$ higher constraining power on $f\sigma_8$ compared to an analysis with $s_{\rm min}=6 \, \hmpc.$ Second, the inclusion of scales smaller than $s\lesssim 1 \, \hmpc$ provides negligible information on $f\sigma_8$ that is not already contained in larger-scale modes. This is true even though the posteriors on some HOD parameters continue to improve as scales $s_{\rm min} <1 \, \hmpc$ are included in the analysis. This indicates that when trying to improve constraining power on $f\sigma_8$ by including smaller scales in an RSD-multipole analysis of LOWZ galaxies, we reach a point of diminishing returns at $s_{\rm min}\lesssim1 \, \hmpc,$ since the bulk of the degeneracies between $f\sigma_8$ and the galaxy--halo connection have already been broken.

Whereas the left-hand panel of Figure~\ref{fig:scale_cuts_precision} illustrates the gain in cosmological constraining power we have achieved by including small-scale information in our analysis, the right-hand panel shows the loss in constraining power that would be incurred by discarding measurements on larger spatial scales. The vertical axis in the right-hand panel shows the strength of the constraints on $f\sigma_8$ derived by an analysis restricted to $400 \, \hkpc < s < s_{\rm max},$ plotted as a function of $s_{\rm max}$ on the horizontal axis. Thus larger values of the horizontal axis correspond to RSD analyses that include more information from large scales, and so each curve decreases monotonically with $s_{\rm max}.$

The salient feature of the right-hand panel of Figure~\ref{fig:scale_cuts_precision} is the shallow slope of each curve. For example, an RSD analysis over the range of scales $s \lesssim 30 \, \hmpc$ has practically the same constraining power as an analysis with $s \lesssim 60 \, \hmpc,$ and an analysis limited to $s_{\rm max}\lesssim 10 \, \hmpc$ only suffers a $50\%$ loss in precision relative to our fiducial all-scale analysis. Considering the left- and right-hand panels of Figure~\ref{fig:scale_cuts_precision} together, we conclude that for RSD analyses of galaxies in the BOSS LOWZ sample, most of the multipole information about $f\sigma_8$ is contained in the scales $2 \, \hmpc \lesssim s \lesssim 20 \, \hmpc$ \citep[for similar conclusions based on a forecasting analysis of cluster-galaxy cross-correlations, see][]{Salcedo2020_MNRAS_491_3061}. Evidently, once the cosmological information content of the quasi-to-nonlinear regime has been harvested, large-scale modes contain only very modest additional information about structure growth.  

This observation has important implications for the computational demands of RSD analyses. Covariance matrix estimation of cosmological observables requires a large number of independently simulated volumes, which even for present-day analyses typically number in the hundreds or thousands in order to sufficiently sample the variance of the largest-scale modes measured in the analysis \citep[e.g.,][]{Hartlap2007_AA_464_399}. Discarding measurements on larger scales and restricting attention to $s_{\rm max}\lesssim 30\ \hmpc$ only mildly degrades the constraints, but reduces the cosmological volume of the simulations needed to generate independent realizations of the predicted data vector. This benefit may prove to be particularly important in the coming decade of cosmological analyses, since the number of independent realizations required by covariance matrix estimation increases sharply with the dimension of the predicted data vector \citep[][]{Taylor2013_MNRAS_432_1928}.

\subsection{Galaxy--halo connection}

In addition to the cosmological growth rate, our analysis also puts constraints on the galaxy--halo connection, marginalised over cosmology. For example, within statistical uncertainties, we find no strong evidence for galaxy assembly bias, i.e. our results are compatible with $A_{\rm cen} = A_{\rm sat} = 0$. However, our constraints are very broad, partially due to the degeneracy between galaxy assembly bias parameters and $f \sigma_8$ \citep{Lange2019_MNRAS_490_1870}. That we do not find evidence for assembly bias from galaxy clustering is in contrast to previous studies by \cite{Zentner2019_MNRAS_485_1196} and \cite{Yuan2020_arXiv_2010_4182}. The former study finds evidence for galaxy assembly bias by fitting $w_{\rm p}$ in the SDSS main galaxy sample and the latter study by analysing the anisotropic two-point correlation function in BOSS CMASS. We note that both galaxy samples are different from the BOSS LOWZ sample. In principle, there can be galaxy assembly bias in both the SDSS and BOSS CMASS samples while being absent in the BOSS LOWZ sample. Similarly, that both studies assume a fixed cosmology might also contribute to their claimed detection of galaxy assembly bias. However, \cite{Yuan2020_arXiv_2010_4182} analyse a total of $\sim 600,000$ galaxies, an order of magnitude larger than the two samples we analyse. Thus, it is well possible that due to the high precision clustering measurements their detection of galaxy assembly bias is robust to marginalisation over cosmology.

In terms of central galaxy velocity bias, we find some evidence for a positive bias in the high-redshift sample, $\alpha_{\rm c} = 0.238_{-0.081}^{+0.082}$. This corresponds to a scenario in which the central galaxy has an additional velocity scatter with respect to the halo core. However, our data cannot firmly rule out $\alpha_{\rm c} = 0$. In contrast, the low-redshift sample prefers such a no velocity bias scenario. Given that the velocity offset is expressed with respect to the halo core, our results of a small to non-existent central velocity bias are in good agreement with other observational studies \citep{Reid2014_MNRAS_444_476, Guo2016_MNRAS_459_3040} and theoretical predictions by \cite{Ye2017_ApJ_841_45}. Unfortunately, a direct comparison with \cite{Guo2015_MNRAS_446_578} and \cite{Guo2015_MNRAS_453_4368} is not possible because in these studies the central velocity bias is expressed with respect to the average velocity of the inner $25\%$ and $100\%$ of dark matter particles, respectively. Generally, this leads to higher inferred values for $\alpha_{\rm c}$ \citep{Ye2017_ApJ_841_45}. When expressed with respect to the halo core, the results of the theoretical study of \cite{Ye2017_ApJ_841_45} indicate $\alpha_c \sim 0.05 - 0.10$ for galaxies with host halo masses $10^{13} - 10^{14} \, h^{-1} M_\odot$, the halo mass range we are probing.

Additionally, our results in both the low and the high-redshift samples are compatible with satellites tracing dark matter spatially, i.e. $\eta = 1$, and having no velocity bias with respect to the prediction from Jeans equilibrium, i.e. $\alpha_{\rm s} = 1$. The result on the spatial bias depends on the stellar mass of the satellite galaxy \citep{Lange2020_arXiv_2011_2377}, making a direct comparison with other works difficult. In the low-redshift, low-stellar mass SDSS main galaxy sample, the data prefers satellites being less concentrated than dark matter \citep[see e.g.][]{Lange2019_MNRAS_487_3112}. However, the situation is less clear for the more massive galaxies in the BOSS CMASS sample where previous studies have obtained good fits to the data without the need for $\eta \neq 1$ \citep{Reid2014_MNRAS_444_476, Guo2015_MNRAS_446_578}. Finally, we obtain no strong constraint on the satellite velocity bias parameter $\alpha_{\rm s}$, contrary to \cite{Guo2015_MNRAS_446_578}. This can be largely attributed to a degeneracy with cosmology that was not accounted for in \cite{Guo2015_MNRAS_446_578} and the smaller galaxy samples analysed here.

Both LOWZ samples target LRGs of comparable number densities and redshifts. Thus, their inferred galaxy-halo connection parameters should be similar but not necessarily identical. We can use this as another consistency check of our results. For example, the high-redshift sample has a lower number density than the low-redshift sample. Thus, we expect the average halo mass to host galaxies to increase, i.e. we expect $M_{\rm min}$ and $M_1$ to be larger for the high-redshit sample. Indeed, this is what we infer, as shown in Table \ref{tab:posterior_galaxy_halo_connection}. Besides these two parameters that are very directly determined by the number density of the sample, all other parameters should be similar between the redshift bins. Reassuringly, as shown in Table \ref{tab:posterior_galaxy_halo_connection}, to within statistical uncertainties these other parameters are identical between the two redshift bins. Finally, the high-redshift sample targets more luminous galaxies. Because centrals are most often the brightest galaxies in their respective haloes, we expect the satellite fraction $f_{\rm sat}$ to decrease with the luminosity of the sample. Our results of $f_{\rm sat} = 14.8 \pm 1.5 \%$ for $z \sim 0.25$ and $f_{\rm sat} = 11.4 \pm 3.0 \%$ for $z \sim 0.4$ are consistent with this expectation.

Finally, we also find that certain galaxy--halo connection parameters are degenerate with the $f \sigma_8$ constraints. The strongest correlations are found for $M_1$ that regulates the number of satellites per halo, $\alpha_{\rm c}$ the amount of central velocity bias and $A_{\rm cen}$ the central velocity bias parameter. This implies that observations providing independent constraints on these parameters could tighten our cosmological constraints in the future.

\subsection{Lensing is low}
\label{subsec:lensing_is_low}

\begin{figure}
    \includegraphics{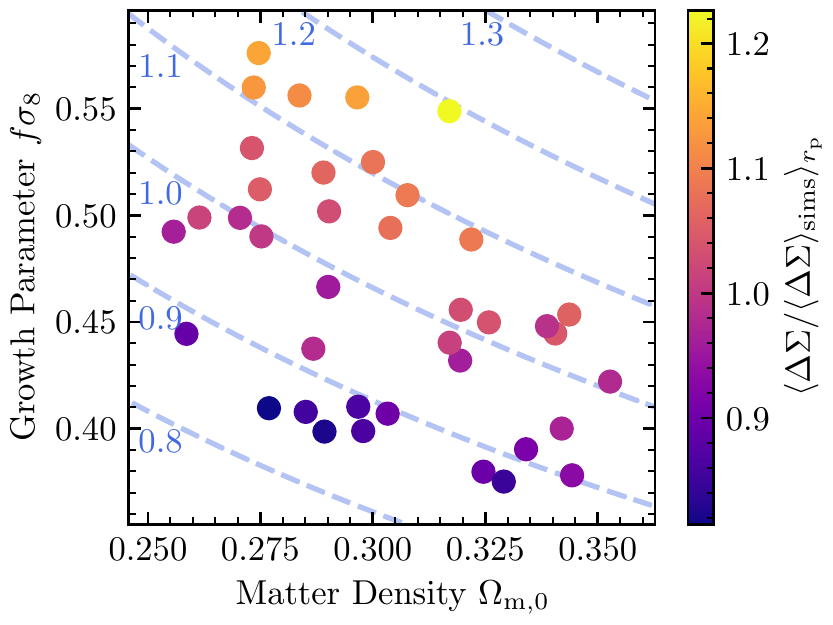}
    \caption{The dependence of the lensing amplitude on cosmological parameter. Results are shown for the $0.18 < z \leq 0.30$ sample. Each dot corresponds to one of the Aemulus simulations. The colour indicates the scale-averaged ratio of the mean posterior lensing amplitude divided by the lensing prediction of marginalisation over cosmology, i.e. the blue line in the left panel of Figure~\ref{fig:lensing_predictions}. Blue lines show a simple power-law fit to the dependence of this ratio on $\Omega_{\rm m}$ and $f \sigma_8$.}
    \label{fig:lensing_cosmology}
\end{figure}

Given Figure \ref{fig:lensing_predictions}, it is interesting to ask what cosmological parameters would alleviate the lensing tension for the low-redshift sample. The galaxy-galaxy lensing amplitude is well-known to be a function of $\Omega_{\rm m, 0}$ and $\sigma_8 (z)$. Thus, in Figure~\ref{fig:lensing_cosmology} we show how the lensing amplitudes from the cosmology-dependent best-fit galaxy--halo models scale with $f \sigma_8$ and $\Omega_{\rm m, 0}$. A simple linear fit reveals that the scale-average lensing amplitude roughly scales with $\Omega_{\rm m, 0}^{0.6} \left(f \sigma_8\right)^{0.9}$. This relation helps explain why the high-redshift sample that predicts a $10\%$ lower $f \sigma_8$ than the CMB $\Lambda$CDM analysis does not overpredict the lensing amplitude, while the low-redshift sample with an inferred $f \sigma_8$ matching the $\Lambda$CDM CMB prediction does. In order to fit the lensing amplitude of the $0.18 < z \lesssim 0.30$ sample, one would likely have to assume a cosmology with a lower matter density or less growth of structure than the best-fit \cite{PlanckCollaboration2020_AA_641_6} $\Lambda$CDM values. Note that while both lensing and RSDs are sensitive to $\sigma_8 (z)$, lensing additionally depends on $\Omega_{\rm m, 0}$ while RSDs has a extra dependence on the growth rate $f$ which is determined by both $\Omega_{\rm m, 0}$ and the dark energy equation of state parameter $w_0$. Thus, a combined lensing and clustering analysis will be sensitive to deviations from $\Lambda$CDM for which $w_0 = -1$. Similarly, gravitational lensing is sensitive to the impact of gravity on relativistic particles whereas RSDs measure its impact on non-relativistic particles. Thus, a lensing plus redshift-space clustering study would also be sensitive to deviations from general relativity. We leave a careful combined cosmological analysis of RSDs and galaxy-galaxy lensing to future work. Such a combined RSD plus lensing analysis would also have to marginalise over complex galaxy assembly bias models and the effect of baryonic feedback.

Finally, \cite{Zu2020_arXiv_2010_1143} recently proposed that the lensing amplitude mismatch can be solved on small scales, $r_{\rm p} < 1 \, \hmpc$, by more complex galaxy--halo models. Among others, the author suggests to assume sub-Poisson distributions for the number of satellites per dark matter halo. This works by suppressing the one-halo galaxy-galaxy clustering terms, particularly the \textit{Fingers of God} effect at fixed HOD. In turn, this allows more satellites for a given clustering amplitude. Since satellites have a lower lensing amplitude at fixed host halo mass than centrals, assuming a sub-Poisson satellite distribution leads to lower predicted lensing amplitudes. However, \cite{Zu2020_arXiv_2010_1143} only considered galaxy--halo models constrained by the projected correlation function $w_{\rm p}$, not the redshift-space correlation function which tightly constrains BOSS satellite fractions. We show in appendix \ref{sec:subpoisson} that the change in the posterior satellite fraction is only at the level of a few percent, even when assuming a maximally sub-Poisson satellite distribution. In turn, this would only lower the lensing amplitude by the same level. Thus, non-Poisson numbers for satellites alone are unlikely to solve the ``lensing is low'' tension on small scales.

\section{Summary and conclusion}
\label{sec:conclusion}

In this work, we perform the first cosmological RSD analysis of BOSS data using simulations with varying cosmology. The simulation-based modelling approach based on the Aemulus suite \citep{DeRose2019_ApJ_875_69} allows us to perform the analysis on scales from $0.4 \, \hmpc$ to $63 \, \hmpc$, significantly extending what is possible with a purely analytic modelling framework. Our analysis uses a sophisticated HOD modelling approach that accounts for galaxy assembly bias and velocity bias. The large range in scales also allows us to place some of the tightest constraints on the growth parameter $f \sigma_8$ recorded to date, even after marginalising over all galaxy--halo connection parameters.

We first test the recovery of $f \sigma_8$ on mock catalogues constructed from the SHAM model of \cite{Lehmann2017_ApJ_834_37}. These mocks contain many of the complexities one would expect from realistic galaxy populations and can alter galaxy clustering substantially, including galaxy assembly bias \citep{Zentner2014_MNRAS_443_3044, McCarthy2019_MNRAS_487_2424, Padilla2019_MNRAS_486_582}, ellipsoidal satellite populations and halo alignment \citep{vanDaalen2012_MNRAS_424_2954}, satellite velocity bias \citep{Guo2015_MNRAS_446_578, Ye2017_ApJ_841_45} as well as non-Poisson satellite numbers \citep{Jiang2017_MNRAS_472_657, Jimenez2019_MNRAS_490_3532}. The mock catalogues are constructed from a $4 \, (h^{-1} \, \mathrm{Gpc})^3$ volume and projected onto the three simulation axes. Thus, the mock measurements have a substantially higher signal-to-noise ratio than the observations which come from a single projection of a volume of $< 1 (h^{-1} \, \mathrm{Gpc})^3$. Despite the complexity in the mock catalogues and the substantially higher precision of the mock measurements compared to observations, we find that the growth rate parameters are accurately recovered to within statistical uncertainties.

We then apply our analysis framework to two roughly volume-limited samples of LRGs at redshifts $z \sim 0.25$ and $z \sim 0.4$. We infer $f \sigma_8 = 0.471 \pm 0.024$ and $0.431 \pm 0.025$ for the low and the high-redshift sample, respectively. The corresponding predictions from the \cite{PlanckCollaboration2020_AA_641_6} CMB analysis under $\Lambda$CDM cosmology are $0.470 \pm 0.006$ and $0.476 \pm 0.005$, respectively. Thus, the low-redshift sample agrees very well with the CMB predictions whereas the high-redshift sample falls $\sim 10\%$ lower at $\sim 2\sigma$ significance. Overall, while our RSD results follow the trend of low-redshift observations suggesting less structure growth than $\Lambda$CDM CMB predictions, when considering both samples together, our RSD results are still in good agreement with the \cite{PlanckCollaboration2020_AA_641_6} $\Lambda$CDM forecasts. We show that most of the cosmological constraining power of the analysis comes from scales $2 \, \hmpc \lesssim s \lesssim 20 \, \hmpc$, i.e. from substantially smaller scales than those commonly analysed in cosmological large-scale structure studies. Thus, despite only analysing a small fraction of the BOSS galaxy sample, our constraints on $f \sigma_8$ are nominally more stringent than those of any analysis focusing on large scales, i.e. $s \gtrsim 30 \hmpc$ only. Overall, in agreement with previous works \citep{Cacciato2013_MNRAS_430_767, Reid2014_MNRAS_444_476, Wibking2019_MNRAS_484_989, Zhai2019_ApJ_874_95, Lange2019_MNRAS_490_1870}, our study highlights the potential benefits of extending the cosmological analysis into the non-linear regime.

We also study the consistency of small and large-scale constraints. We find 1-2.5$\sigma$ hints that our analysis prefers smaller $f \sigma_8$ when analysing small scales only ($s \lesssim 10 \, \hmpc$) compared to large scales only ($s \gtrsim 6 \, \hmpc$). The difference is the strongest for the low-redshift sample, reaching roughly $2.5 \sigma$ significance. However, our model is able to fit observations on all scales well with $\chi_\nu^2 \approx 1$ and we do not observe a similar scale dependence in the much more accurate mock observations. New data from DESI will yield more constraining measurements and will be able to determine if the difference in $f \sigma_8$ constraints is a statistical fluke, a sign of modelling systematics, or some form of new physics. We stress the importance of conducting more mock tests in the future to test the veracity of cosmological constraints from all-scale analyses such as the one performed here. Ideally, the mock tests would be based on hydrodynamical simulations or semi-analytic models of galaxy formation. Such tests will be critical for upcoming studies with higher precision measurements.

In the future, we plan to include galaxy-galaxy lensing in our all-scale cosmological analysis. In this work, we have shown how our predictions for the galaxy-galaxy lensing amplitude compares against measurements from SDSS \citep{Singh2020_MNRAS_491_51}. We find that the lensing amplitude is underpredicted for the low-redshift sample and in good agreement for the high-redshift sample. As expected, the lensing amplitude at fixed clustering is positively correlated $\Omega_{\rm m, 0}$ and $\sigma_8 (z)$. This likely contributes to the high-redshift sample with lower implied $f \sigma_8$ showing a better agreement with the measured lensing amplitude. Overall, a combined lensing and redshift-space clustering analysis would likely find evidence for less growth of structure or lower matter density $\Omega_{\rm m, 0}$ than implied by the CMB plus $\Lambda$CDM forecast. However, a full cosmological interpretation of the lensing amplitude would involve marginalising over baryonic feedback \citep{Leauthaud2017_MNRAS_467_3024, Lange2019_MNRAS_490_1870} as well as more tests on mock catalogues. Additionally, one could combine galaxy-galaxy lensing measurements with observations of the Sunyaev–Zeldovich effect to directly constraint the strength of baryonic feedback \citep{Schaan2020_arXiv_2009_5557, Amodeo2020_arXiv_2009_5558}. We leave such a detailed study to future work but note that it could provide further interesting measurements on the growth of structure in the late Universe, possible violations to general relativity and the validity of the $\Lambda$CDM model.

\section*{Acknowledgements}

We thank the Aemulus collaboration for making their simulations publicly available and for the letting us take part in the Aemulus mock challenge that motivated the mock tests presented in this study. We also thank Sukhdeep Singh for providing the SDSS galaxy-galaxy lensing measurements, Sihan Yuan for useful conversations on RSD modelling as well as Kuan Wang and Risa Wechsler for commenting on earlier versions of this manuscript.

We acknowledge use of the lux supercomputer at UC Santa Cruz, funded by NSF MRI grant AST 1828315. This material is based on work supported by the U.D Department of Energy, Office of Science, Office of High Energy Physics under Award Number DE-SC0019301. AL acknowledges support from the David and Lucille Packard foundation, and from the Alfred P. Sloan foundation. HG acknowledges the support from the National Natural Science Foundation of China (Nos. 11833005, 11922305). Work done by APH was supported by the U.S. Department of Energy, Office of Science, Office of Nuclear Physics, under contract DE-AC02-06CH11357.

This work made use of the following software packages: {\sc matplotlib} \citep{Hunter2007_CSE_9_90}, {\sc SciPy}, {\sc NumPy} \citep{vanderWalt2011_CSE_13_22}, {\sc Astropy} \citep{AstropyCollaboration2013_AA_558_33}, {\sc Colossus} \citep{Diemer2015_ascl_soft_1016}, {\sc halotools} \citep{Hearin2017_AJ_154_190}, {\sc MultiNest} \citep{Feroz2008_MNRAS_384_449, Feroz2009_MNRAS_398_1601, Feroz2019_OJAp_2_10}, {\sc PyMultiNest} \citep{Buchner2014_AA_564_125}, {\sc scikit-learn} \citep{Pedregosa2012_arXiv_1201_0490}, {\sc Spyder} and {\sc GNOME \LaTeX}.

This research used resources of the National Energy Research Scientific Computing Center (NERSC), a U.S. Department of Energy Office of Science User Facility located at Lawrence Berkeley National Laboratory, operated under Contract No. DE-AC02-05CH11231.

Funding for SDSS-III has been provided by the Alfred P. Sloan Foundation, the Participating Institutions, the National Science Foundation, and the U.S. Department of Energy Office of Science. The SDSS-III web site is \url{http://www.sdss3.org/}.

SDSS-III is managed by the Astrophysical Research Consortium for the Participating Institutions of the SDSS-III Collaboration including the University of Arizona, the Brazilian Participation Group, Brookhaven National Laboratory, Carnegie Mellon University, University of Florida, the French Participation Group, the German Participation Group, Harvard University, the Instituto de Astrofisica de Canarias, the Michigan State/Notre Dame/JINA Participation Group, Johns Hopkins University, Lawrence Berkeley National Laboratory, Max Planck Institute for Astrophysics, Max Planck Institute for Extraterrestrial Physics, New Mexico State University, New York University, Ohio State University, Pennsylvania State University, University of Portsmouth, Princeton University, the Spanish Participation Group, University of Tokyo, University of Utah, Vanderbilt University, University of Virginia, University of Washington, and Yale University.

\section*{Data Availability}

The Aemulus and UNIT simulations used in this article are publicly available at \url{https://aemulusproject.github.io/} and \url{http://www.unitsims.org/}, respectively. The SDSS data analysed is available at \url{https://www.sdss.org/}. All derived data generated in this research as well as code used will be shared on reasonable request to the corresponding author.

\bibliographystyle{mnras}
\bibliography{bibliography}

\appendix

\section{Simulation statistics}
\label{section:simulation_statistics}

Tables \ref{tab:evidence} and \ref{tab:evidence_test} show the cosmological parameters, goodness-of-fit values and evidence values with respect to the two BOSS LOWZ samples assuming the default analysis choices.

\begin{table*}
    \centering
    \begin{tabular}{c|c|c|c|c|c|c|c|ccc|ccc}
    \multirow{2}{*}{Name} & $H_0$ & \multirow{2}{*}{$w_0$} & \multirow{2}{*}{$\sigma_8$} & \multirow{2}{*}{$\Omega_{m, 0}$} & \multirow{2}{*}{$\Omega_{b, 0}$} & \multirow{2}{*}{$n_s$} & \multirow{2}{*}{$N_{\rm eff}$} & \multicolumn{3}{c|}{z=0.25} & \multicolumn{3}{c}{z=0.40} \\
    & $[\mathrm{km} \, \mathrm{s}^{-1} \, \mathrm{Mpc}^{-1}]$ &&&&&&& $f \sigma_8$ & $\Delta\ln\mathcal{Z}$ & $\chi^2$ & $f \sigma_8$ & $\Delta\ln\mathcal{Z}$ & $\chi^2$ \\
    \hline
    B00 & 63.4 & -0.82 & 0.77 & 0.34 & 0.056 & 0.98 & 2.92 & 0.445 & -5.0 & 33.3 & 0.444 & -3.3 & 30.4 \\
    B01 & 73.1 & -1.13 & 0.90 & 0.26 & 0.042 & 0.98 & 3.17 & 0.499 & -1.1 & 23.0 & 0.517 & -5.7 & 31.5 \\
    B02 & 63.7 & -0.68 & 0.69 & 0.32 & 0.057 & 1.00 & 3.26 & 0.380 & -12.0 & 44.4 & 0.379 & -5.1 & 27.3 \\
    B03 & 64.0 & -0.74 & 0.67 & 0.33 & 0.055 & 0.95 & 3.56 & 0.375 & -7.6 & 36.9 & 0.375 & -8.5 & 39.1 \\
    B04 & 65.0 & -0.77 & 0.75 & 0.30 & 0.052 & 0.97 & 2.66 & 0.407 & -2.8 & 26.9 & 0.410 & -3.7 & 28.2 \\
    B05 & 72.8 & -1.33 & 0.93 & 0.28 & 0.039 & 0.93 & 2.96 & 0.556 & -7.1 & 30.9 & 0.573 & -23.4 & 61.3 \\
    B06 & 62.7 & -0.71 & 0.71 & 0.34 & 0.058 & 0.97 & 2.71 & 0.400 & -11.3 & 42.9 & 0.398 & -4.1 & 27.8 \\
    B07 & 64.4 & -0.87 & 0.78 & 0.34 & 0.055 & 0.97 & 3.94 & 0.453 & -2.6 & 27.3 & 0.454 & -2.6 & 27.3 \\
    B08 & 69.4 & -1.16 & 0.89 & 0.30 & 0.043 & 0.95 & 3.60 & 0.525 & -5.6 & 27.6 & 0.536 & -12.5 & 41.4 \\
    B09 & 62.4 & -0.83 & 0.72 & 0.35 & 0.055 & 0.95 & 3.90 & 0.422 & -2.8 & 28.4 & 0.421 & -2.9 & 25.1 \\
    B10 & 72.1 & -1.24 & 0.85 & 0.29 & 0.042 & 0.96 & 4.24 & 0.502 & -2.6 & 27.1 & 0.515 & -11.4 & 42.7 \\
    B11 & 67.7 & -0.86 & 0.81 & 0.29 & 0.049 & 1.00 & 2.83 & 0.437 & -1.9 & 25.1 & 0.445 & -3.9 & 28.0 \\
    B12 & 65.4 & -0.88 & 0.79 & 0.33 & 0.053 & 0.95 & 2.88 & 0.450 & -2.3 & 26.5 & 0.452 & -1.5 & 23.3 \\
    B13 & 71.1 & -1.12 & 0.87 & 0.28 & 0.043 & 0.98 & 3.00 & 0.490 & -1.0 & 23.5 & 0.505 & -6.4 & 34.7 \\
    B14 & 68.7 & -1.12 & 0.92 & 0.32 & 0.048 & 0.97 & 2.75 & 0.549 & -10.6 & 35.7 & 0.557 & -14.7 & 42.9 \\
    B15 & 74.1 & -1.30 & 0.91 & 0.27 & 0.039 & 0.93 & 3.73 & 0.532 & -3.6 & 26.0 & 0.549 & -17.7 & 53.6 \\
    B16 & 70.1 & -1.13 & 0.81 & 0.29 & 0.044 & 0.97 & 3.77 & 0.466 & -1.8 & 25.8 & 0.478 & -4.7 & 29.2 \\
    B17 & 74.4 & -1.25 & 0.87 & 0.26 & 0.040 & 0.95 & 3.22 & 0.492 & -1.2 & 24.3 & 0.512 & -7.0 & 35.7 \\
    B18 & 70.8 & -1.03 & 0.74 & 0.28 & 0.046 & 0.95 & 4.28 & 0.410 & -2.7 & 26.5 & 0.420 & -7.1 & 34.9 \\
    B19 & 72.4 & -1.09 & 0.81 & 0.26 & 0.043 & 0.97 & 3.68 & 0.444 & -0.3 & 23.1 & 0.460 & -3.7 & 28.5 \\
    B20 & 67.1 & -0.99 & 0.84 & 0.32 & 0.049 & 0.95 & 3.39 & 0.489 & -1.1 & 22.9 & 0.494 & -5.7 & 31.2 \\
    B21 & 66.4 & -0.87 & 0.76 & 0.32 & 0.054 & 0.98 & 3.85 & 0.432 & -3.0 & 28.0 & 0.435 & -2.5 & 28.9 \\
    B22 & 68.1 & -1.03 & 0.88 & 0.31 & 0.046 & 0.96 & 2.62 & 0.509 & -1.6 & 24.4 & 0.518 & -10.6 & 40.8 \\
    B23 & 62.0 & -0.57 & 0.57 & 0.32 & 0.059 & 0.97 & 3.47 & 0.305 & -34.6 & 78.4 & 0.303 & -31.7 & 74.2 \\
    B24 & 63.0 & -0.76 & 0.69 & 0.33 & 0.057 & 0.96 & 4.15 & 0.390 & -4.2 & 31.3 & 0.390 & -4.2 & 27.5 \\
    B25 & 65.7 & -0.95 & 0.79 & 0.32 & 0.048 & 0.93 & 3.09 & 0.456 & -1.7 & 25.8 & 0.460 & -3.6 & 29.0 \\
    B26 & 71.8 & -1.13 & 0.90 & 0.27 & 0.043 & 0.94 & 2.79 & 0.512 & -0.7 & 20.9 & 0.527 & -12.1 & 42.8 \\
    B27 & 67.4 & -0.96 & 0.73 & 0.30 & 0.047 & 0.97 & 4.02 & 0.410 & -2.4 & 26.8 & 0.417 & 0.0 & 21.5 \\
    B28 & 74.8 & -1.40 & 0.96 & 0.27 & 0.039 & 0.96 & 3.81 & 0.576 & -14.0 & 43.6 & 0.596 & -23.3 & 59.6 \\
    B29 & 71.4 & -1.24 & 0.93 & 0.30 & 0.044 & 0.94 & 3.43 & 0.555 & -7.5 & 31.0 & 0.568 & -18.5 & 50.5 \\
    B30 & 73.4 & -1.22 & 0.87 & 0.27 & 0.041 & 0.96 & 4.07 & 0.499 & 0.0 & 22.3 & 0.516 & -9.9 & 40.8 \\
    B31 & 73.8 & -1.38 & 0.94 & 0.27 & 0.039 & 0.96 & 3.34 & 0.560 & -9.2 & 35.3 & 0.579 & -15.0 & 45.7 \\
    B32 & 68.4 & -0.93 & 0.72 & 0.29 & 0.048 & 0.95 & 3.98 & 0.399 & -2.2 & 26.6 & 0.406 & -4.0 & 23.5 \\
    B33 & 66.1 & -0.88 & 0.78 & 0.32 & 0.054 & 0.99 & 3.64 & 0.440 & -1.7 & 26.0 & 0.444 & -2.3 & 27.0 \\
    B34 & 69.1 & -1.03 & 0.86 & 0.30 & 0.048 & 0.95 & 3.13 & 0.494 & -2.2 & 25.8 & 0.503 & -6.5 & 32.1 \\
    B35 & 61.7 & -0.61 & 0.68 & 0.34 & 0.062 & 1.00 & 3.05 & 0.378 & -12.6 & 49.4 & 0.374 & -4.9 & 27.1 \\
    B36 & 70.4 & -1.11 & 0.91 & 0.29 & 0.044 & 0.97 & 3.30 & 0.520 & -2.7 & 24.2 & 0.533 & -14.8 & 47.9 \\
    B37 & 66.7 & -0.85 & 0.73 & 0.30 & 0.051 & 0.98 & 3.51 & 0.399 & -4.7 & 31.9 & 0.404 & -6.7 & 32.9 \\
    B38 & 69.7 & -0.96 & 0.74 & 0.29 & 0.049 & 0.98 & 4.11 & 0.408 & -5.0 & 31.9 & 0.416 & -2.8 & 27.6 \\
    B39 & 64.7 & -0.94 & 0.76 & 0.34 & 0.052 & 0.96 & 4.19 & 0.448 & -3.6 & 30.4 & 0.450 & -3.4 & 29.1 \\
    \end{tabular}
    \caption{Overview of the simulations used in this work and their corresponding evidence and goodness-of-fit values with respect to the observations.}
    \label{tab:evidence}
\end{table*}

\begin{table*}
    \centering
    \begin{tabular}{c|c|c|c|c|c|c|c|ccc|ccc}
    \multirow{2}{*}{Name} & $H_0$ & \multirow{2}{*}{$w_0$} & \multirow{2}{*}{$\sigma_8$} & \multirow{2}{*}{$\Omega_{m, 0}$} & \multirow{2}{*}{$\Omega_{b, 0}$} & \multirow{2}{*}{$n_s$} & \multirow{2}{*}{$N_{\rm eff}$} & \multicolumn{3}{c|}{z=0.25} & \multicolumn{3}{c}{z=0.40} \\
    & $[\mathrm{km} \, \mathrm{s}^{-1} \, \mathrm{Mpc}^{-1}]$ &&&&&&& $f \sigma_8$ & $\Delta\ln\mathcal{Z}$ & $\chi^2$ & $f \sigma_8$ & $\Delta\ln\mathcal{Z}$ & $\chi^2$ \\
    \hline
    T00-0 & 63.2 & -0.73 & 0.69 & 0.33 & 0.058 & 0.98 & 2.95 & 0.386 & -7.2 & 35.9 & 0.386 & -1.9 & 25.8 \\
    T00-1 & 63.2 & -0.73 & 0.69 & 0.33 & 0.058 & 0.98 & 2.95 & 0.386 & -7.9 & 37.0 & 0.386 & -5.4 & 27.6 \\
    T00-2 & 63.2 & -0.73 & 0.69 & 0.33 & 0.058 & 0.98 & 2.95 & 0.386 & -10.1 & 43.2 & 0.386 & -3.0 & 30.0 \\
    T00-3 & 63.2 & -0.73 & 0.69 & 0.33 & 0.058 & 0.98 & 2.95 & 0.386 & -12.6 & 47.6 & 0.386 & -6.5 & 32.9 \\
    T00-4 & 63.2 & -0.73 & 0.69 & 0.33 & 0.058 & 0.98 & 2.95 & 0.386 & -9.8 & 41.4 & 0.386 & -5.8 & 33.8 \\
    T01-0 & 65.7 & -0.86 & 0.75 & 0.31 & 0.053 & 0.97 & 3.20 & 0.424 & -3.0 & 29.5 & 0.428 & -3.0 & 25.4 \\
    T01-1 & 65.7 & -0.86 & 0.75 & 0.31 & 0.053 & 0.97 & 3.20 & 0.424 & -3.5 & 28.4 & 0.428 & -2.5 & 26.2 \\
    T01-2 & 65.7 & -0.86 & 0.75 & 0.31 & 0.053 & 0.97 & 3.20 & 0.424 & -3.3 & 29.3 & 0.428 & -2.2 & 25.5 \\
    T01-3 & 65.7 & -0.86 & 0.75 & 0.31 & 0.053 & 0.97 & 3.20 & 0.424 & -3.9 & 28.9 & 0.428 & -1.8 & 26.4 \\
    T01-4 & 65.7 & -0.86 & 0.75 & 0.31 & 0.053 & 0.97 & 3.20 & 0.424 & -2.9 & 28.8 & 0.428 & -4.9 & 27.0 \\
    T02-0 & 68.2 & -1.00 & 0.81 & 0.30 & 0.048 & 0.96 & 3.45 & 0.461 & -1.6 & 25.4 & 0.469 & -3.1 & 27.3 \\
    T02-1 & 68.2 & -1.00 & 0.81 & 0.30 & 0.048 & 0.96 & 3.45 & 0.461 & -1.2 & 25.6 & 0.469 & -3.2 & 28.3 \\
    T02-2 & 68.2 & -1.00 & 0.81 & 0.30 & 0.048 & 0.96 & 3.45 & 0.461 & -1.1 & 25.3 & 0.469 & -5.0 & 30.0 \\
    T02-3 & 68.2 & -1.00 & 0.81 & 0.30 & 0.048 & 0.96 & 3.45 & 0.461 & -3.5 & 28.4 & 0.469 & -1.8 & 23.7 \\
    T02-4 & 68.2 & -1.00 & 0.81 & 0.30 & 0.048 & 0.96 & 3.45 & 0.461 & -1.1 & 24.6 & 0.469 & -5.3 & 32.3 \\
    T03-0 & 70.7 & -1.13 & 0.86 & 0.29 & 0.043 & 0.95 & 3.70 & 0.496 & -0.6 & 22.6 & 0.508 & -8.6 & 38.4 \\
    T03-1 & 70.7 & -1.13 & 0.86 & 0.29 & 0.043 & 0.95 & 3.70 & 0.496 & -0.7 & 23.4 & 0.508 & -5.2 & 31.9 \\
    T03-2 & 70.7 & -1.13 & 0.86 & 0.29 & 0.043 & 0.95 & 3.70 & 0.496 & 0.2 & 21.5 & 0.508 & -5.6 & 29.9 \\
    T03-3 & 70.7 & -1.13 & 0.86 & 0.29 & 0.043 & 0.95 & 3.70 & 0.496 & -1.5 & 24.8 & 0.508 & -4.3 & 30.0 \\
    T03-4 & 70.7 & -1.13 & 0.86 & 0.29 & 0.043 & 0.95 & 3.70 & 0.496 & -1.9 & 24.1 & 0.508 & -10.6 & 40.3 \\
    T04-0 & 73.2 & -1.27 & 0.91 & 0.28 & 0.040 & 0.94 & 3.95 & 0.531 & -5.5 & 27.3 & 0.548 & -18.7 & 49.5 \\
    T04-1 & 73.2 & -1.27 & 0.91 & 0.28 & 0.040 & 0.94 & 3.95 & 0.531 & -3.2 & 24.8 & 0.548 & -11.8 & 39.6 \\
    T04-2 & 73.2 & -1.27 & 0.91 & 0.28 & 0.040 & 0.94 & 3.95 & 0.531 & -6.9 & 31.6 & 0.548 & -17.0 & 50.2 \\
    T04-3 & 73.2 & -1.27 & 0.91 & 0.28 & 0.040 & 0.94 & 3.95 & 0.531 & -2.1 & 24.2 & 0.548 & -9.5 & 38.2 \\
    T04-4 & 73.2 & -1.27 & 0.91 & 0.28 & 0.040 & 0.94 & 3.95 & 0.531 & -3.3 & 25.7 & 0.548 & -14.4 & 47.4 \\
    T05-0 & 69.7 & -1.09 & 0.82 & 0.28 & 0.045 & 0.95 & 3.70 & 0.464 & 1.2 & 20.7 & 0.476 & -3.9 & 28.0 \\
    T05-1 & 69.7 & -1.09 & 0.82 & 0.28 & 0.045 & 0.95 & 3.70 & 0.464 & -0.9 & 23.3 & 0.476 & -8.8 & 37.9 \\
    T05-2 & 69.7 & -1.09 & 0.82 & 0.28 & 0.045 & 0.95 & 3.70 & 0.464 & -0.3 & 22.1 & 0.476 & -6.2 & 33.6 \\
    T05-3 & 69.7 & -1.09 & 0.82 & 0.28 & 0.045 & 0.95 & 3.70 & 0.464 & -1.6 & 26.2 & 0.476 & -4.0 & 29.0 \\
    T05-4 & 69.7 & -1.09 & 0.82 & 0.28 & 0.045 & 0.95 & 3.70 & 0.464 & -0.3 & 23.0 & 0.476 & -4.6 & 33.6 \\
    T06-0 & 66.7 & -0.90 & 0.80 & 0.32 & 0.051 & 0.97 & 3.20 & 0.457 & -2.5 & 26.9 & 0.461 & -8.3 & 38.4 \\
    T06-1 & 66.7 & -0.90 & 0.80 & 0.32 & 0.051 & 0.97 & 3.20 & 0.457 & -5.0 & 32.7 & 0.461 & -5.9 & 34.0 \\
    T06-2 & 66.7 & -0.90 & 0.80 & 0.32 & 0.051 & 0.97 & 3.20 & 0.457 & -3.6 & 30.5 & 0.461 & -6.1 & 33.6 \\
    T06-3 & 66.7 & -0.90 & 0.80 & 0.32 & 0.051 & 0.97 & 3.20 & 0.457 & -4.8 & 30.2 & 0.461 & -1.5 & 22.0 \\
    T06-4 & 66.7 & -0.90 & 0.80 & 0.32 & 0.051 & 0.97 & 3.20 & 0.457 & -2.6 & 29.6 & 0.461 & -2.2 & 26.3 \\
    \end{tabular}
    \caption{Similar to Table~\ref{tab:evidence} but for the test simulations of the Aemulus simulation suite.}
    \label{tab:evidence_test}
\end{table*}

\section{Scatter in cosmological evidence}

When building a model for the cosmological evidence $\mathcal{Z} (\mathcal{C} | \mathbf{D})$, we make the simplifying assumption that it depends solely on $f \sigma_8$ and, at most, one additional cosmological parameter. At the same time, it is evident from Figures \ref{fig:log_ev_fs8_challenge}, \ref{fig:log_ev_fs8_w0_challenge} and \ref{fig:log_ev_fs8} that the models for $\mathcal{Z} (\mathcal{C} | \mathbf{D})$ does not fit the data perfectly. A concern is that scatter between calculated and model evidence values is due to unmodelled cosmological parameter dependencies and that neglecting such dependencies biases our marginal constraints on $f \sigma_8$.

First, not all unmodelled dependencies necessarily bias the constraints on $f \sigma_8$. For example, if the dependence on a second cosmological parameter $\gamma$ is perfectly separable, i.e. $\mathcal{Z} (f \sigma_8, \gamma | \mathbf{D}) = f(f \sigma_8) g(\gamma)$, neglecting the dependence on $\gamma$ should not significantly bias the constraint on $f \sigma_8$. Additionally, we have performed systematic checks in section \ref{sec:mock_tests} that show that our results are robust with respect to various modelling choices.

However, we can use the test simulations of the Aemulus simulation suite to investigate more directly what the source of the scatter is. The test simulations of the Aemulus suite are $7$ groups of $5$ simulations, each. All simulations of the same group share the exact same cosmological parameters but different random seeds for the initial conditions. Thus, any scatter in the evidence in a group of simulations with the same cosmological parameters is entirely due to random fluctuations. In Table~\ref{tab:evidence_test}, we show the cosmological evidence for all test simulations with respect to our two observational data sets. For each set of $5$ simulations, we estimate the variance in the scatter,
\begin{equation}
    \hat{S}^2_{\ln \mathcal{Z}} = \frac{1}{5 - 1} \sum\limits_{i=1}^5 \left[ \ln \mathcal{Z}_i (\mathcal{C} | \mathbf{D}) - \mathrm{Mean} (\ln \mathcal{Z} (\mathcal{C} | \mathbf{D})) \right]^2.
\end{equation}
For the $7 \times 2 = 14$ combinations of simulation sets and observations, the average variance in the scatter is $\langle \hat{S}^2_{\ln \mathcal{Z}} \rangle = 3.8$. At the same time, when fitting a model to $\mathcal{Z} (\mathcal{C} | \mathbf{D})$, we also obtain an estimate for the scatter $S^2_{\ln \mathcal{Z}}$ that is due to both statistical fluctuations and possibly unmodelled parameters. If the scatter in the evidence values around the best-fit were primarily due to unmodelled cosmological parameter dependencies, we would overestimate the scatter found for the test simulations where the only source of scatter is statistical fluctuations. However, our model predicts $\langle \hat{S}^2_{\ln \mathcal{Z}} \rangle$ to be in the range $2.0 - 5.9$ ($95\%$ certainty), agreeing well with our finding of $3.8$. Thus, the observed scatter in the evidence values of the test simulations are compatible with cosmic variance being the major source of scatter in Figures \ref{fig:log_ev_fs8_challenge}, \ref{fig:log_ev_fs8_w0_challenge} and \ref{fig:log_ev_fs8}.

\section{Sub-Poisson scatter for satellites}
\label{sec:subpoisson}

\begin{figure*}
    \centering
    \subfloat{\includegraphics{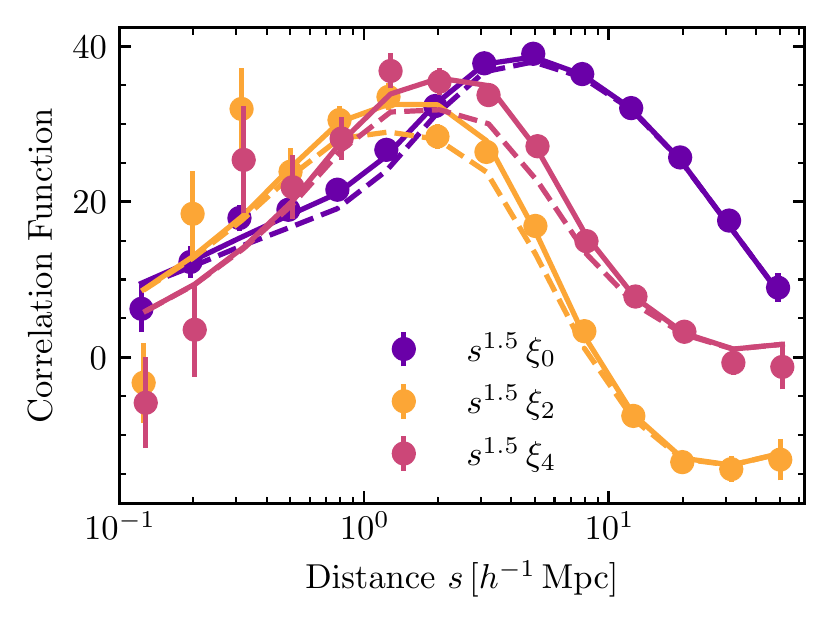}}
    \subfloat{\includegraphics{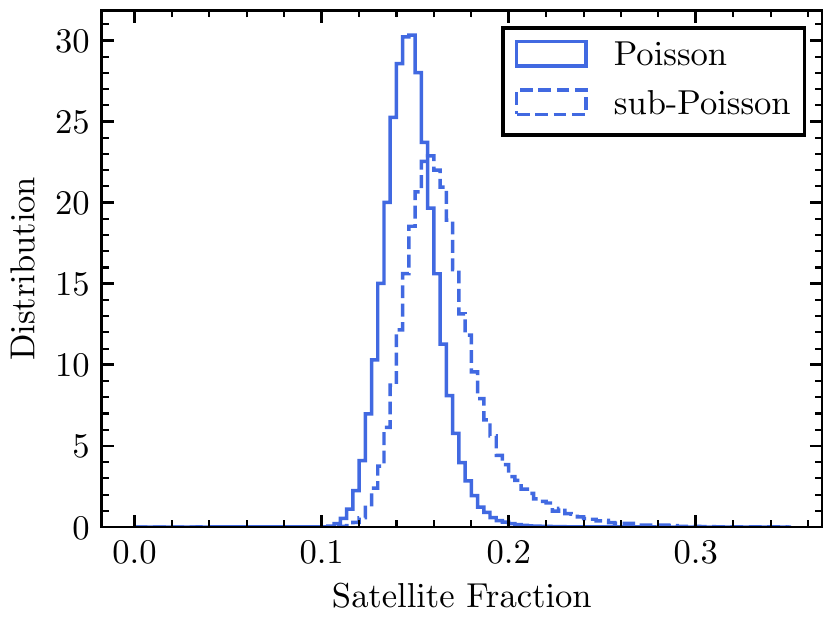}}
    \caption{Impact of the second moment of the satellite distribution on observables (left) and posterior constraints on the satellite fraction (right). In both panels, we compare models assuming a Poisson distribution for satellites (solid) against a maximally sub-Poisson distribution (dashed).}
    \label{fig:subpoisson}
\end{figure*}

In our default analysis, we assume that the satellite occupation numbers at fixed halo mass and concentration follow a Poisson distribution. This is motivated by the finding that subhaloes that host satellite galaxies follow a distribution that is close to Poisson \citep{Jiang2017_MNRAS_472_657}. Recently, \cite{Zu2020_arXiv_2010_1143} has advocated for maximally sub-Poisson satellite numbers in order to alleviate the so-called ``lensing is low'' tension. This integer distribution with the smallest possible scatter assigns $i$ satellites with a probability of $p = 1 - \langle N_s \rangle + i$ and $i + 1$ satellites with a probability $1 - p$, where $i$ is an integer with $i \leq \langle N_s \rangle < i + 1$. For this distribution, one can show $\langle N_s (N_s - 1) \rangle = 2 \langle N_s \rangle i - i (i + 1)$.

We note that subhalo occupation numbers, while slightly sub-Poisson for low occupation numbers, are far from maximally sub-Poisson \citep{Jiang2017_MNRAS_472_657}. Additionally, whether each subhalo host a satellite is determined by galaxy formation physics, an additional stochastic process. If galaxies hosted by subhaloes evolve independently of each other, the distribution will be driven towards a Poisson distribution. Similarly, if satellite properties inside each halo are are positively correlated with each other, an effect called 1-halo galactic conformity \citep{Weinmann2006_MNRAS_366_2}, this will drive the distribution of satellites further towards a super-Poisson distribution. In both cases, one would expect the satellite number distribution to be even further away from a maximally sub-Poisson distribution than subhaloes.

Nonetheless, we have performed the analysis for the $0.18 \leq z < 0.30$ LOWZ sample assuming a maximally sub-Poisson distribution for satellites. In Figure~\ref{fig:subpoisson}, we show how a change in the satellite number distribution affects the observables. The galaxy--halo model for both the Poisson and sub-Poisson model is the best-fit model for the $0.18 \leq z < 0.30$ LOWZ sample when Poisson numbers are assumed. Thus, all differences arise entirely from changes to the second moment of the satellite distribution. For the sub-Poisson distribution, we observe that all the multipoles of the redshift-space correlation function are slightly suppressed on small scales. This makes sense since assuming sub-Poisson satellite number suppresses the 1-halo satellite-satellite terms which scales with $\langle N_s (N_s - 1) \rangle$. Although $\langle N_s (N_s - 1) \rangle$ is strongly suppressed compared to the Poisson assumption, the effect is weak because the 1-halo central-satellite term dominates over the 1-halo satellite-satellite terms \citep[see e.g.][]{Zheng2016_MNRAS_458_4015} and is unaffected by the second moment of the satellite distribution. 

Given the significant changes in the multipoles at fixed cosmology and galaxy--halo connection parameters, it is evident that assumption on the satellite numbers will affect posterior constraints. For example, the right panel of Figure~\ref{fig:subpoisson} demonstrates how cosmology-averaged inferences on the satellite fraction $f_{\rm sat}$ are affected by the choice of satellite distribution. Assuming a maximally sub-Poisson distribution implies slightly higher satellite fractions. This makes sense since a sub-Poisson distribution reduces the strength of the \textit{Fingers of God} effect and the latter is positively correlated with $f_{\rm sat}$. However, even in the extreme case of a maximally sub-Poisson distribution, the shift in the posterior constraint on $f_{\rm sat}$ compared to the Poisson scenario is very small, $\Delta f_{\rm sat} \sim 0.02$. Thus, by itself, assuming a strongly sub-Poisson satellite distribution is likely not enough to make the large satellite fractions advocated for in \cite{Zu2020_arXiv_2010_1143} compatible with RSD constraints in this work or other results from the literature \citep{Reid2014_MNRAS_444_476, Guo2015_MNRAS_446_578, Saito2016_MNRAS_460_1457}.

Finally, we do not find that the choice of the second moment for the satellite distribution impacts the cosmological constraint significantly: Assuming a maximally sub-Poisson distribution for satellites reduces the inferred $f \sigma_8$ posterior by only $0.012$ or around $0.5 \sigma$. More modest variations from Poisson distributions would likely result in even smaller shifts of $f \sigma_8$.

\label{lastpage}

\end{document}